\documentclass[prd,twocolumn,eqsecnum,nofootinbib]{revtex4}
\pdfoutput=1
\usepackage{bm} 
\usepackage{amssymb}
\usepackage{graphicx} 
\DeclareFontFamily{OT1}{rsfs}{} 
\DeclareFontShape{OT1}{rsfs}{m}{n}{<-7> rsfs5 
    <7-10> rsfs7 <10-> rsfs10}{}   
\DeclareMathAlphabet{\scr}{OT1}{rsfs}{m}{n}
\newcommand{\E}{{\cal E}} 
\newcommand{\B}{{\cal B}} 
\newcommand{\stf}[1]{{\langle {#1} \rangle}}
\newcommand{\U}[1]{{\mbox{}_{\scriptscriptstyle #1} U}} 
\newcommand{\V}[1]{{\mbox{}_{\scriptscriptstyle #1} V}} 
\newcommand{\PPsi}[1]{{\mbox{}_{\scriptscriptstyle #1}\! \Psi}} 
\begin{document}
\title{Nonrotating black hole in a post-Newtonian tidal environment} 
\author{Stephanne Taylor and Eric Poisson}
\affiliation{Department of Physics, University of Guelph, Guelph, 
Ontario, Canada N1G 2W1}
\date{September 11, 2008} 
\begin{abstract} 
We examine the motion and tidal dynamics of a nonrotating black hole
placed within a post-Newtonian external spacetime. The black hole's
gravity is described accurately to all orders in $Gm/c^2 r$, where $m$
is the black-hole mass and $r$ is the distance to the black hole. The
tidal perturbation created by the external environment is treated as a
small perturbation. At a large distance from the black hole, the
gravitational field of the external distribution of matter is assumed
to be sufficiently weak to be adequately described by the (first)
post-Newtonian approximation to general relativity. There, the black 
hole is treated as a monopole contribution to the total gravitational
field. There exists an overlap in the domains of validity of each 
description, and the black-hole and post-Newtonian metrics are matched  
in the overlap. The matching procedure produces (i) a
justification of the statement that a nonrotating black hole is a
post-Newtonian monopole; (ii) a complete characterization of the
coordinate transformation between the inertial, barycentric frame and
the accelerated, black-hole frame; (iii) the equations of motion for
the black hole; and (iv) the gravito-electric and gravito-magnetic
tidal fields acting on the black hole. We first calculate the
equations of motion and tidal fields by making no assumptions
regarding the nature of the post-Newtonian environment; this could  
contain a continuous distribution of matter (so as to model a galactic
core) or any number of condensed bodies. We next specialize our
discussion to a situation in which the black hole is a member of a
post-Newtonian two-body system. As an application of our results, we
examine the geometry of the deformed event horizon and calculate the
tidal heating of the black hole, the rate at which it acquires mass as
a result of its tidal interaction with the companion body.   
\end{abstract} 
\maketitle

\section{Introduction and summary} 

\subsection{This work and its context} 

How does a black hole move in an external spacetime, and what effects 
do the tidal fields created in the external spacetime have on the
black hole? These are the questions that are investigated in this
work, in a context in which the black hole is nonrotating and the
gravity of the external universe is sufficiently weak to be adequately
described by the post-Newtonian approximation to general
relativity. This work is a continuation of a line of inquiry that was
initiated by Manasse \cite{manasse:63} in the early nineteen sixties,
and that has been pursued to the present day.  

To pose our questions more precisely, and to better discuss the place
of this work in the context of what was achieved previously, we
introduce two length scales that are relevant to this problem. The
first is set by $m$, the mass of the black hole, which gives rise to
an associated length scale $M := Gm/c^2$, the gravitational radius of
the black hole. The second is ${\cal R}$, the radius of curvature of
the external spacetime, evaluated at the black hole's position. Our
work, and all others that preceded it, is carried out in a context in
which $M/{\cal R} \ll 1$, so that there is a clean separation between
these scales. Only in this context can one meaningfully speak of
a black hole moving in an external spacetime; when $M$ is comparable
to ${\cal R}$, no distinction can be made between the ``black hole''
and the ``external spacetime.'' 

As a concrete example we may consider a situation in which the black
hole is a member of a binary system. Then 
${\cal R} \sim \sqrt{b^3/M_{\rm tot}}$, where $b$ is the separation
between the bodies and $M_{\rm tot} := G(m+m')/c^2$ is a measure of
the total mass within the system ($m'$ is the external mass). In this
case we have  
\[
\frac{M}{\cal R} \sim \frac{M}{M_{\rm tot}} 
\Bigl( \frac{M_{\rm tot}}{b} \Bigr)^{3/2}, 
\]
and for this work, this is required to be small. There are two
particular ways to achieve this. In the {\it small-hole
approximation} the black-hole mass is assumed to be much smaller than
the external mass, so that $M/M_{\rm tot} \sim m/m' \ll 1$; then 
$M/{\cal R}$ is small irrespective of the size of $M_{\rm tot}/b$, and
the binary system can be strongly relativistic. In the {\it weak-field 
approximation} it is $M_{\rm tot}/b$ that is assumed to be small,
while the mass ratio is left unconstrained.  

Our work is concerned with the weak-field approximation. The black
hole is placed within a post-Newtonian external spacetime, and the
external gravitational potentials determine its motion as well as the
tidal gravity acting upon it. We determine the motion of the black 
hole, the tidal fields, and the effects of the tidal fields on the
structure of spacetime around the black hole, all within the
post-Newtonian approximation to general relativity. At first, we do
not specify the nature of the post-Newtonian environment. We leave it  
completely general; the black hole might be immersed within a smooth
distribution of matter (a model for a galactic core, for example), or
it might be part of an $N$-body system (with the number, nature, and
state of motion of the bodies left arbitrary). As our
work progresses, we specialize our results to a two-body system
undergoing generic orbital motion, and finally we examine the special
case of circular orbits.  

The motion of a black hole in an arbitrary external spacetime was
first investigated by D'Eath \cite{death:75a, death:75b, death:96} and
Kates \cite{kates:80}, who showed that in the limit $M/{\cal R} \to
0$, the black hole moves on a geodesic of the external spacetime. In
this limit the black hole behaves as a test mass, in spite of the fact
that the self-gravity of the black hole never ceases to be strong. The 
corrections to geodesic motion produced by the coupling of the
black-hole spin with the curvature of the external spacetime were
worked out by Thorne and Hartle \cite{thorne-hartle:85}, who also
obtained precession equations for the spin vector. These authors
exploited the power of matched asymptotic expansions in their
derivation of the equations of motion. In their approach, the metric
of the black hole (deformed by the conditions in the external
spacetime) is matched to the metric of the external spacetime
(perturbed by the moving black hole). The matching is carried out in a
region in which both descriptions are valid, and it produces both the
equations of motion and the tidal fields, with only the Einstein field
equations as additional input. Our work is a continuation of this
program. 

These investigations were next specialized to systems for which the
gravity of the external spacetime is weak; this is the context that
interests us in this paper. Demianski and Grishchuk
\cite{demianski-grishchuk:74} showed that to leading order in a
post-Newtonian expansion of the external gravity, the black hole moves 
according to the Newtonian equations of motion. Their results were
generalized to first post-Newtonian order by D'Eath \cite{death:75b}
and Damour \cite{damour:83}, who found agreement between the equations
of motion for black holes in binary systems and the standard
(Einstein-Infeld-Hoffman) equations of motion of post-Newtonian
theory. Our work is a continuation of this effort, and our results are  
slightly more general than theirs: While our black hole is still
immersed within a post-Newtonian environment, this environment is
completely general, and the black hole is not required to be a member
of a binary system. When, however, we specialize our results to this
particular case, we recover the results of D'Eath and Damour.   

The motion of a black hole in a post-Newtonian external spacetime is
well understood, and our contribution to this understanding is a
relatively minor one. The same cannot be said, however, of the effects
of the external tidal gravity on the black hole, which have not been
much discussed in the literature. This is the true focus of this work,
and our main goal in this paper is to calculate the post-Newtonian
tidal fields acting on the black hole, and to explore the physical
consequences of the tidal interaction.  

We are not claiming that ours is the first calculation of
post-Newtonian tidal fields acting on a self-gravitating body. It is
not. In their pioneering work on relativistic celestial mechanics,
Damour, Soffel, and Xu   
\cite{damour-soffel-xu:91, damour-soffel-xu:92, damour-soffel-xu:93}
calculated the post-Newtonian tidal fields acting on an
arbitrarily-structured body with weak internal gravity. This work was
recently generalized to arbitrarily-structured, strongly-gravitating
bodies by Racine and Flanagan \cite{racine-flanagan:05}. Our work is
concerned instead with a very specific type of strongly
self-gravitating body: a nonrotating black hole. We calculate the
post-Newtonian tidal fields acting on this black hole, and observe
that they are the same as those obtained by Damour, Soffel, and Xu in
the case of weakly self-gravitating monopoles. We confirm, therefore,
the general expectation (known as the ``effacement principle'') that
the post-Newtonian tidal fields must depend on the body's multipole
moments only (in addition to the conditions in the external
spacetime), and not on additional details concerning its internal
structure.  

The effects of tidal fields on the structure of spacetime around a
black hole were first investigated by Manasse \cite{manasse:63}, who
provided an essential input to the work later carried out by D'Eath,
Kates, Thorne, and Hartle. Adopting the small-hole approximation
defined previously, Manasse calculated the metric around a small black  
hole that falls radially toward a much larger black hole. Each black
hole was taken to be nonrotating, and the small hole was taken to move
on a geodesic of the (unperturbed) Schwarzschild spacetime of the
large hole. The tidal gravity exerted by the large black hole 
creates a perturbation in the Schwarzschild metric of the
small hole, and employing the techniques of Regge and Wheeler
\cite{regge-wheeler:57}, Manasse was able to calculate this
perturbation in a local neighborhood of the black hole. His metric is
expressed as an expansion in powers of $r/{\cal R}$, where $r$ is
the distance to the small hole; it is accurate through second order in
$r/{\cal R}$, and it is valid to all orders in $M/r$, which measures
the strength of the small hole's self-gravity. The case of circular
motion around a large Schwarzschild black hole was treated much later
by Poisson \cite{poisson:04a}.  

The methods used by Manasse are not necessarily restricted to the
small-hole approximation. Alvi \cite{alvi:00, alvi:03}
realized that these methods could be seamlessly extended to the
general setting defined by the requirement $M/{\cal R} \ll 1$, which
includes both the small-hole and weak-field approximations as special
cases. Alvi exploited this insight to calculate the tidal fields
acting on a black hole in a post-Newtonian binary system (perhaps
with another black hole). In Alvi's work, the two bodies have
comparable masses and the black hole
has a significant influence on the geometry of the external
spacetime. Alvi calculated the tidal fields to leading (Newtonian)
order in the post-Newtonian approximation to general relativity, for 
circular orbits. The metric of the distorted black hole was next
joined to the post-Newtonian two-body metric, and the global metric
was presented in a single coordinate system that corotates with the
system. (In Alvi's original work there is a discontinuity in the
metric at the common boundary between the two descriptions. The joint
was made continuous in a follow-up paper by Yunes {\it et al.}
\cite{yunes-etal:06}.)     

Alvi's insight was exploited by Poisson \cite{poisson:05} in a
calculation of the metric of a nonrotating black hole placed within an
arbitrary tidal environment (still restricted by $M/{\cal R} \ll
1$). Working, like Manasse \cite{manasse:63}, in a local neighborhood
of the black hole, Poisson was able to calculate the metric through
third order in $r/{\cal R}$, while keeping the expressions accurate to
all orders in $M/r$. Poisson's metric is parameterized by a number of
{\it tidal moments}, freely-specifiable tensorial functions of time
that characterize the black hole's tidal environment. This metric
gives a general description of the spacetime around a black hole in
{\it any} tidal environment, but a more complete description requires
the determination of the tidal moments. This is our task in this
paper: We calculate the tidal moments for the specific case described
above, in which the black hole is immersed within a post-Newtonian
external spacetime. This is a generalization of Alvi's work
\cite{alvi:00, alvi:03, yunes-etal:06}: We appeal to the weak-field 
approximation, we calculate the tidal fields created by an arbitrary  
post-Newtonian spacetime, and we do so to a higher order of accuracy
than what was achieved by Alvi. 

\subsection{Our results} 

The metric of a nonrotating black hole immersed in a tidal environment
is expressed as a perturbation of the Schwarzschild metric. We take
the black hole to have a mass $m$ when it is in complete isolation
(unperturbed), and we denote its gravitational radius by $M :=
Gm/c^2$. The strength of the tidal perturbation is measured by the
inverse length scale ${\cal R}^{-1}$, and we assume that $M/{\cal R}
\ll 1$; the tidal perturbation is weak. In addition, we assume that
the black hole moves in an empty region of spacetime, so that in the
hole's neighborhood $\scr B$, the
perturbed metric satisfies the vacuum field equations linearized about 
the exact Schwarzschild solution. We present the metric in the
comoving reference frame of the black hole, in a quasi-Cartesian
system of coordinates $\bar{x}^\alpha = (\bar{x}^0, \bar{x}^a) =
(c\bar{t}, \bar{x}, \bar{y}, \bar{z})$ that enforce the harmonic
conditions $\partial_\beta (\sqrt{-g} g^{\alpha\beta}) = 0$. Our
convention is that Greek indices run from 0 to 3, while Latin indices
cover the spatial coordinates and run from 1 to 3. We raise and lower
Latin indices with the Euclidean metric $\delta_{ab}$, and we let 
$\epsilon_{abc}$ denote the permutation symbol of ordinary vector
calculus (with $\epsilon_{123} = \epsilon_{xyz} = 1$). 

The time-time and time-space components of the black-hole metric are
(Sec.~III) 
\begin{eqnarray} 
g_{\bar{0}\bar{0}} &=& -\frac{1-M/\bar{r}}{1+M/\bar{r}} 
- \frac{1}{c^2} (1-M/\bar{r})^2 \bar{\cal E}_{ab}(\bar{t}) 
  \bar{x}^a \bar{x}^b 
\nonumber \\ & & 
+ O(\bar{r}^3/{\cal R}^3), 
\label{1.1} \\
g_{\bar{0}\bar{a}} &=& \frac{2}{3c^3} (1-M/\bar{r})(1+M/\bar{r})^2   
  \epsilon_{abp} \bar{\cal B}^p_{\ c}(\bar{t}) 
  \bar{x}^b \bar{x}^c 
\nonumber \\ & & 
+ O(\bar{r}^3/{\cal R}^3); 
\label{1.2} 
\end{eqnarray} 
for our purposes here we shall not need an expression for
$g_{\bar{a}\bar{b}}$, the space-space components of the metric. The
metric is expressed as an expansion in powers of $\bar{r}/{\cal R}$,
the ratio of $\bar{r} := \sqrt{\bar{x}^2+\bar{y}^2+\bar{z}^2}$, the
distance from the black hole, to ${\cal R}$, the local radius of
curvature of the external spacetime. The metric is valid in the black
hole's local neighborhood $\scr B$, which is defined by $\bar{r} <
\bar{r}_{\rm max}$ (see Fig.~1); we demand that 
$\bar{r}_{\rm max}/{\cal R}$ be small, but within $\scr B$ the ratio
$M/\bar{r}$ is allowed to be arbitrarily large.   

\begin{figure} 
\includegraphics[width=0.9\linewidth]{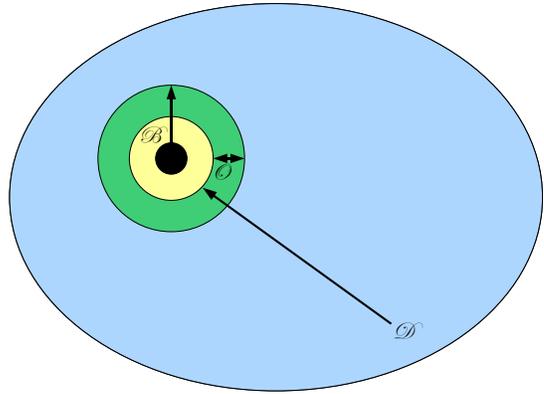}
\caption{The post-Newtonian domain $\scr D$, the black-hole
  neighborhood $\scr B$, and the overlap region $\scr O$. The
  post-Newtonian domain is depicted in blue (light grey), and it
  includes the green (dark grey) annulus that surrounds the black
  hole. The black-hole neighborhood is drawn as a yellow (white) disk 
  around the black hole, and it also includes the green (dark grey)
  annulus. The black hole is represented as a black disk. The overlap
  region is the union of ${\scr B}$ and ${\scr D}$ that is shown in
  green (dark grey), the union of the blue and yellow colors.}   
\end{figure} 

The first term on the right-hand side of Eq.~(\ref{1.1}) is the
Schwarzschild piece of the metric; it is expressed in harmonic
coordinates, and in these coordinates the (unperturbed) event horizon
is situated at $\bar{r} = M$. The second term and the right-hand side
of Eq.~(\ref{1.2}) represent the tidal perturbation. The tensorial
functions $\bar{\cal E}_{ab}(\bar{t})$ and  
$\bar{\cal B}_{ab}(\bar{t})$ are the {\it tidal moments} (Sec.~II),
and it is these tensors that characterize the black hole's tidal 
environment. The tidal moments are symmetric and tracefree (STF)
tensors, in the sense that $\bar{\cal E}_{ba} = \bar{\cal E}_{ab}$
and $\delta^{ab} \bar{\cal E}_{ab} = 0$, with similar relations
holding for $\bar{\cal B}_{ab}$. The tensors $c^{-2} \bar{\E}_{ab}$
and $c^{-3} \bar{\B}_{ab}$ have a dimension of squared inverse length,
and their scale {\it defines} ${\cal R}$, the local radius of
curvature; we have $c^{-2} \bar{\cal E}_{ab} \sim
{\cal R}^{-2}$ and $c^{-3} \bar{\cal B}_{ab} \sim {\cal R}^{-2}$. The
tidal moments are not determined by solving the Einstein field
equations in $\scr B$. They are {\it a priori} arbitrary, and their
determination is accomplished by matching the black-hole metric to a
global metric defined in a domain that is much larger than 
${\scr B}$. In this work the tidal moments are determined by placing
the black hole within a post-Newtonian environment;
the global metric is obtained by solving the Einstein field equations
in the first post-Newtonian approximation.    

We next describe this post-Newtonian environment. We consider a
spatial domain ${\scr D}$ that is much larger than ${\scr B}$, the
black-hole neighborhood (see Fig.~1). This domain contains an
arbitrary distribution of matter\footnote{Here and below, the word
``matter'' describes a number of different situations. The matter   
could be a continuous fluid, so as to model an accretion disk or a
galactic core. It could also correspond to a collection of $N-1$
bodies with weak self-gravity, making the black hole a member of an
$N$-body system. Or else the domain $\scr D$ could exclude a number
$N-1$ of small regions that would each contain a condensed body such
as a neutron star or a black hole. In this last case, the
post-Newtonian domain would contain no matter at all, but we will
nevertheless refer to the $N-1$ excluded regions as ``matter.''}, and
it is assumed that everywhere within ${\scr D}$, gravity is
sufficiently weak to be adequately described by the post-Newtonian
approximation to general relativity. The domain is spatially limited
by a sphere of radius $r_{\rm near}$ centered on the post-Newtonian
barycenter. This sphere marks the boundary of the near zone: If 
$\cal T$ is a typical time scale for processes taking place within 
${\scr D}$, and if $\lambda_c = c {\cal T}$ is a typical wavelength
of the gravitational waves escaping the domain, then
$r_{\rm near} < \lambda_c$. The domain also excludes a sphere of
radius $\bar{r}_{\rm min}$ centered on the black hole, inside which
gravity is too strong to be adequately described by post-Newtonian
theory. We demand both that $M/\bar{r}_{\rm min} \ll 1$ and 
$\bar{r}_{\rm min}/{\cal R} \ll 1$, which is possible when $M \ll
{\cal R}$. There exists an overlap region ${\scr O}$ between the
black-hole neighborhood ${\scr B}$ and the post-Newtonian domain 
$\scr D$. This region is described by
$\bar{r}_{\rm min} < \bar{r} < \bar{r}_{\rm max}$, and we assume that 
there is no matter in ${\scr O}$. So while matter is present
somewhere within ${\scr D}$, we assume that the black hole is moving
in an empty region of spacetime.  

The metric in ${\scr D}$ can be expressed as a post-Newtonian
expansion of the form (Sec.~IV) 
\begin{eqnarray} 
g_{00} &=& -1 + \frac{2}{c^2} U + \frac{2}{c^4} (\Psi - U^2) 
+ O(c^{-6}), 
\label{1.3} \\ 
g_{0a} &=& -\frac{4}{c^3} U_a + O(c^{-5}), 
\label{1.4} \\
g_{ab} &=& \biggl( 1 + \frac{2}{c^2} U \biggr) \delta_{ab} 
+ O(c^{-4}), 
\label{1.5}
\end{eqnarray} 
in which $U$ is a Newtonian potential, $U_a$ a vector potential, and
$\Psi$ a post-Newtonian potential; the metric is presented in harmonic
coordinates $x^\alpha = (x^0, x^a) = (ct, x, y, z)$. These barycentric
coordinates differ from the black-hole coordinates
$(c\bar{t},\bar{x},\bar{y},\bar{z})$ introduced previously; the
black-hole and post-Newtonian metrics are presented in different
coordinate systems. In spite of the fact that each system is harmonic,
the coordinates are indeed distinct: They are defined in different
domains ($x^\alpha$ in ${\scr D}$, $\bar{x}^\alpha$ in ${\scr B}$),
and they have a different spatial origin ($x^\alpha$ is centered on
the post-Newtonian barycenter, whose position is fixed in the global
reference frame, while $\bar{x}^\alpha$ is centered on the moving
black hole).   

In the overlap region ${\scr O}$ the spacetime is empty of matter, and 
the potentials $U$, $U_a$, and $\Psi$ satisfy the vacuum field
equations of post-Newtonian theory. The Newtonian potential, for
example, must satisfy Laplace's equation in flat space, $\nabla^2 U 
= 0$. The solution must account for the presence of a black hole, and
it must also account for the presence of matter outside ${\scr O}$. We
treat the black hole as a post-Newtonian monopole, and we write   
\begin{equation} 
U(t,\bm{x}) = \frac{Gm}{|\bm{x}-\bm{z}(t)|} + U_{\rm ext}(t,\bm{x}), 
\label{1.6}
\end{equation} 
in which the three-dimensional vector $\bm{z}(t)$ denotes the position
of the black hole in the barycentric coordinates. The external
potential $U_{\rm ext}$ is created by the matter outside ${\scr O}$,
and within the overlap region we have $\nabla^2 U_{\rm ext} = 0$. The 
potentials $U_a$ and $\Psi$ are handled in a similar fashion
(Sec.~IV), and in this way we construct the post-Newtonian metric in
${\scr O}$.   

The post-Newtonian metric of Eqs.~(\ref{1.3})--(\ref{1.5}) and the
black-hole metric of Eqs.~(\ref{1.1})--(\ref{1.2}) both give a valid
description of the gravitational field in ${\scr O}$. The metrics must
agree in the overlap region, and matching them determines the
equations of motion for $\bm{z}(t)$ as well as the tidal moments 
$\bar{\cal E}_{ab}(\bar{t})$ and $\bar{\cal B}_{ab}(\bar{t})$. This
matching, however, can only be done after the post-Newtonian metric is 
transformed from the barycentric coordinates $x^\alpha$ to the
black-hole coordinates $\bar{x}^\alpha$. This transformation, between
two systems of harmonic coordinates, can be fully worked out (Sec.~V),
relying on previous work by Kopeikin \cite{kopeikin:88}, Brumberg and
Kopeikin \cite{brumberg-kopeikin:89}, Damour, Soffel, and Xu
\cite{damour-soffel-xu:91}, Kopeikin and Vlasov
\cite{kopeikin-vlasov:04}, and Racine and Flanagan
\cite{racine-flanagan:05}.  

The matching procedure determines the coordinate transformation
completely (Sec.~VI C), and it produces a justification of the earlier
statement that the black hole can be treated as a post-Newtonian
monopole. (A fuller discussion of this point is presented at the end
of Sec.~IV.) This statement, therefore, is a strict consequence of the
Einstein field equations, rather than an artificial assumption. At
first post-Newtonian order, the gravitational field of a black hole is
that of a pure monopole, and it would be inconsistent to endow the
black hole with an additional multipole structure. 

The matching procedure produces also an equation of motion for the
moving black hole. It reads (Sec.~VI E)   
\begin{eqnarray} 
a^a &=& \partial^a U_{\rm ext} 
+ \frac{1}{c^2} \biggl[ \partial^a \Psi_{\rm ext} 
- 4 \bigl( \partial^a U^b_{\rm ext} - \partial^b U^a_{\rm ext} \bigr)
  v_b 
\nonumber \\ & & 
+ 4 \partial_t U^a_{\rm ext} 
+ \bigl( v^2 - 4 U_{\rm ext} \bigr) \partial^a U_{\rm ext} 
\nonumber \\ & & 
- v^a \bigl( 4 v^b \partial_b U_{\rm ext} 
+ 3 \partial_t U_{\rm ext} \bigr) \biggl] + O(c^{-4}),  
\label{1.7}
\end{eqnarray}
in which $\bm{v} = d\bm{z}/dt$ is the black hole's velocity vector in
the barycentric frame, and $\bm{a} = d\bm{v}/dt$ is its
acceleration. The external potentials $U_{\rm ext}$, $U^a_{\rm ext}$,
and $\Psi_{\rm ext}$ are defined as in Eq.~(\ref{1.6}), and they are
evaluated at $\bm{x} = \bm{z}(t)$ after differentiation. Equation
(\ref{1.7}) applies to a black hole moving in any post-Newtonian
environment. When this environment consists of $(N-1)$ external
bodies, so that the black hole is a member of an $N$-body system, 
Eq.~(\ref{1.7}) reduces to the standard (Einstein-Infeld-Hoffman)
post-Newtonian equations of motion. These are listed, for example, in
Exercise 39.15 of Misner, Thorne, and Wheeler \cite{MTW:73}. In
effect, Eq.~(\ref{1.7}) states that the black hole moves on a
geodesic of the metric of Eqs.~(\ref{1.3})--(\ref{1.5}), in which the 
(singular) potentials $U$, $U_a$, and $\Psi$ are replaced by the
(smooth) external potentials created by the distribution of matter
outside the black-hole neighborhood ${\scr B}$. 

It is interesting to compare the differences between our derivation of 
Eq.~(\ref{1.7}) and the approach followed by Racine and Flanagan (RF) 
\cite{racine-flanagan:05}. First, the work of RF is concerned with
arbitrarily structured bodies (with weak or strong internal gravity),
while our own work is concerned specifically with a nonrotating black
hole, which is necessarily treated as a post-Newtonian monopole. Our
work, therefore, is a specialization of theirs. Second, RF define the 
frame $(\bar{t},\bar{x}^a)$, which they call the body-adapted frame,
by (essentially) setting the body's intrinsic mass dipole moment to
zero; this is (essentially) the piece of $g_{\bar{0}\bar{0}}$ that
behaves as $\bar{x}^a/\bar{r}^{3}$. This coordinate choice does not, 
in general, constrain the tidal dipole moment; this is
(essentially) the piece of $g_{\bar{0}\bar{0}}$ that grows as
$\bar{x}^a$. In our work, the coordinates $(\bar{t},\bar{x}^a)$ are
defined so as to eliminate all mass dipole moments (both intrinsic and
tidal) from the metric. This is made possible by the fact that we are
dealing here with a specific type of body --- a nonrotating black hole
--- instead of a general body whose nature is characterized only by an
infinite set of multipole moments. Indeed, the work of Zerilli
\cite{zerilli:70} shows that in vacuum, an even-parity dipole
perturbation of the Schwarzschild metric can always be removed by a
gauge transformation; it is this gauge choice that defines our own 
version of the body-adapted frame, and the metric of Eqs.~(\ref{1.1}) 
and (\ref{1.2}) reflects the complete absence of dipole terms. Third, 
in RF, the equations of motion are obtained by exploiting the integral
form of the momentum-conservation identities that come as a
consequence of the Landau-Lifshitz formulation of the Einstein field
equations \cite{landau-lifshitz:b2}. In our approach, the equations of
motion are obtained directly by matching the black-hole and
post-Newtonian metrics, and the computations are considerably
simpler. This advantage is intimately tied to our complete control
over the dipole terms; a derivation of the equations of motion
involving matching only would not be possible without the ability 
to set both the intrinsic and tidal mass dipole moments to zero.      

Finally, the matching procedure produces expressions for the tidal
moments (Secs.~VI D and E). In the barycentric frame they are given by  
\begin{widetext}
\begin{eqnarray} 
{\cal E}_{ab} &=& -\partial_{ab} U_{\rm ext} + \frac{1}{c^2} 
\left\lgroup 
-\partial_\stf{ab} \Psi_{\rm ext} 
+ 4v^c \bigl( \partial_{ab} U^{\rm ext}_c 
- \partial_{c\langle a} U^{\rm ext}_{b\rangle} \bigr) 
- 4 \partial_{t\langle a} U^{\rm ext}_{b\rangle} 
- 2(v^2 - U_{\rm ext}) \partial_{ab} U_{\rm ext} 
\right. \nonumber \\ & & \mbox{} \left.
+ 3 v^c v_{\langle a} \partial_{b\rangle c} U_{\rm ext} 
+ 2 v_{\langle a} \partial_{b\rangle t} U_{\rm ext} 
+ 3 \partial_{\langle a} U_{\rm ext} \partial_{b\rangle} U_{\rm ext} 
\right\rgroup + O(c^{-4}) 
\label{1.8}
\end{eqnarray}
\end{widetext}
and 
\begin{equation} 
{\cal B}_{ab} = 2 \epsilon_{pq(a} \partial^p_{\ b)} \bigl( 
U^q_{\rm ext} - v^q U_{\rm ext} \bigr) + O(c^{-2}), 
\label{1.9}
\end{equation} 
in which the external potentials are evaluated at $\bm{x} = \bm{z}(t)$
after differentiation. The brackets around indices indicate
symmetrization, while angular brackets indicate an STF operation: For
any tensor $A_{ab}$ we have $A_{(ab)} = \frac{1}{2}(A_{ab} + A_{ba})$
and $A_\stf{ab} := A_{(ab)} - \frac{1}{3} \delta_{ab} A$, where $A :=
\delta^{ab} A_{ab}$. The expressions of Eqs.~(\ref{1.8}) and
(\ref{1.9}) are valid for any post-Newtonian environment. 

Given the vast difference in notations and ways of expressing our 
results, we have not attempted to compare Eqs.~(\ref{1.8}) and
(\ref{1.9}) to the results obtained by Racine and Flanagan
\cite{racine-flanagan:05}, nor to those of Damour, Soffel, and Xu 
\cite{damour-soffel-xu:91, damour-soffel-xu:92, 
damour-soffel-xu:93}. We can state, however, that the specialization
of Eqs.~(\ref{1.8}) and (\ref{1.9}) to an $N$-body system is in
perfect agreement with the corresponding results of Damour, Soffel,
and Xu --- see, in particular, Eqs.~(4.29)--(4.31) of
Ref.~\cite{damour-soffel-xu:93}. We shall provide evidence for this 
statement in Sec.~VII A. 

When the black hole is part of a binary system in circular motion, the
nonvanishing components of the tidal moments are given by (Sec.~VII C) 
\begin{eqnarray} 
{\cal E}_{11} + {\cal E}_{22} &=& -\frac{Gm'}{b^3} 
\biggl[ 1 + \frac{m}{2(m+m')} (v_{\rm rel}/c)^2 
\nonumber \\ & & \mbox{} 
+ O(c^{-4}) \biggr],  
\label{1.10} \\ 
{\cal E}_{11} - {\cal E}_{22} &=& -\frac{3Gm'}{b^3} 
\biggl[ 1 - \frac{3m+4m'}{2(m+m')} (v_{\rm rel}/c)^2 
\nonumber \\ & & \mbox{} 
+ O(c^{-4}) \biggr] \cos 2\omega t, 
\label{1.11} \\ 
{\cal E}_{12} &=& -\frac{3Gm'}{2b^3} 
\biggl[ 1 - \frac{3m+4m'}{2(m+m')} (v_{\rm rel}/c)^2 
\nonumber \\ & & \mbox{} 
+ O(c^{-4}) \biggr] \sin 2\omega t, 
\label{1.12} \\ 
{\cal B}_{13} &=& -\frac{3Gm'}{b^3} v_{\rm rel} \cos\omega t 
+ O(c^{-2}),  
\label{1.13} \\ 
{\cal B}_{23} &=& -\frac{3Gm'}{b^3} v_{\rm rel} \sin\omega t 
+ O(c^{-2}), 
\label{1.14}
\end{eqnarray} 
where $m$ is the mass of the black hole, $m'$ the mass of the
companion, and $b := |\bm{z}-\bm{z'}|$ the orbital separation between
the two bodies (in barycentric harmonic coordinates); the
components ${\cal E}_{13}$, ${\cal E}_{23}$, ${\cal B}_{11}$, 
${\cal B}_{22}$, and ${\cal B}_{12}$ all vanish for circular
orbits, and ${\cal E}_{33} = -({\cal E}_{11} + {\cal E}_{22})$. 
In these equations $v_{\rm rel} := |\bm{v}-\bm{v'}| =
\sqrt{G(m+m')/b}$ stands for the (Newtonian) orbital velocity of the
relative orbit, and  
\begin{eqnarray} 
\omega &=& \sqrt{\frac{G(m+m')}{b^3}} \Bigl[
1 - \frac{3m^2 + 5mm' + 3m^{\prime 2}}{2(m+m')^2} (v_{\rm rel}/c)^2   
\nonumber \\ & & \mbox{} 
+ O(c^{-4}) \Bigr]  
\label{1.15}
\end{eqnarray}
is the orbit's post-Newtonian angular velocity. Equations
(\ref{1.8})--(\ref{1.15}) are expressed in the barycentric
coordinates $x^\alpha$. In the main text (see Sec.~VII D) we also list
expressions that are valid in the black hole's moving frame. In the
comoving coordinates $\bar{x}^\alpha$ we find that
Eqs.~(\ref{1.10})--(\ref{1.14}) are unchanged, except for the fact
that the tidal moments must now be expressed in terms of the
transformed phase variable $\bar{\omega} \bar{t}$. The
transformed angular velocity $\bar{\omega}$, given by 
\begin{eqnarray} 
\bar{\omega} &=& \sqrt{\frac{G(m+m')}{b^3}} \Bigl[
1 - \frac{3m^2 + 7mm' + 3m^{\prime 2}}{2(m+m')^2} (v_{\rm rel}/c)^2  
\nonumber \\ & & \mbox{} 
+ O(c^{-4}) \Bigr],   
\label{1.16}
\end{eqnarray}
accounts for the change in time coordinate as well as the geodetic
precession of the moving frame relative to the barycentric frame. 
Notice the different coefficient in front of $mm'$ in the
post-Newtonian term. 

Equations (\ref{1.7})--(\ref{1.16}) are the main results of this
work. As an application we examine (Sec.~VIII), in a suitable
choice of gauge, the intrinsic geometry of the event horizon of a
tidally-deformed black hole. We also calculate (Sec.~IX A) the rate at
which the black hole acquires mass as a result of its tidal
interaction with the companion body. We find that this rate of tidal
heating is given by  
\begin{eqnarray} 
G\dot{m} &=& \frac{32}{5 c^{15}} \frac{m^6 m^{\prime 2}}{(m+m')^8} 
\Bigl(\frac{G(m+m')}{b}\Bigr)^9 \biggl[ 1 
\nonumber \\ & & \mbox{} 
- \frac{5m^2 + 12mm' + 6m^{\prime 2}}{(m+m')^2} (v_{\rm rel}/c)^2  
\nonumber \\ & & \mbox{} 
+ O(c^{-4}) \biggr]. 
\label{1.17} 
\end{eqnarray}
The rate at which the tidal coupling increases the black hole's
angular momentum $J$ can next be obtained from the rigid-rotation
relation $\dot{m} c^2 = \bar{\omega} \dot{J}$. 

\subsection{Relevance of this work} 

Our interest in this paper is mostly in the exploration of the tidal 
dynamics of black holes, in a weak-field context in which the tidal
fields can be determined explicitly. These dynamics produce the tidal 
heating of the black hole, an increase in mass (and angular momentum
and surface area) that is produced entirely by an influx of
gravitational energy across the horizon. This fascinating phenomenon
was studied before, most notably by Poisson and Sasaki
\cite{poisson-sasaki:95}, Alvi \cite{alvi:01}, Price and Whelan
\cite{price-whelan:01}, Hughes \cite{hughes:01}, Martel
\cite{martel:04}, and Poisson \cite{poisson:04d, poisson:05}. We
provide here some additional insights.  

The tidal heating of a nonrotating black hole is generally very
small. Relative to the energy radiated away by gravitational waves,
the effect is of order $(v/c)^8$. In practical terms, the effect is
likely to be too small to be observed in a gravitational-wave signal
that would be measured by ground-based detectors such as LIGO, VIRGO,
and GEO600. For example, Alvi \cite{alvi:01} calculated that for
binary systems involving black holes with masses ranging from 5 to 50
solar masses, tidal heating is negligible: It contributes only a small 
fraction of a wave cycle during the signal's sweep through the
detector's frequency band.    

In some circumstances, however, the tidal heating is a significant
effect that should not be neglected \cite{price-whelan:01}. 
In particular, it is likely to be observed in gravitational-wave
signals that would be measured by a space-based detector such as
LISA. For example, Martel \cite{martel:04} showed that during a close 
encounter between a massive black hole and a compact body of much
smaller mass, up to approximately five percent of the total radiated
energy is absorbed by the black hole, the rest making its way out
to infinity. Hughes \cite{hughes:01} calculated that when the massive
black hole is rapidly rotating, tidal heating slows down the inspiral
of the orbiting body, thereby increasing the duration of the
gravitational-wave signal. For example, a $1\ M_\odot$ compact body on
a slightly inclined, circular orbit around a $10^6\ M_\odot$ black
hole of near-maximum spin will spend approximately two years in the
LISA frequency band before its final plunge into the black hole;
Hughes shows that tidal heating contributes approximately 20 days
(and $10^4$ wave cycles) to these two years.

Another situation in which the tidal heating might be ``measured'' is
in the numerical simulations of black-hole mergers. To reveal this
small effect the simulations will need to be performed with very high
accuracy and resolution, but this should become possible within the
next few years. The simulations would reveal a steady growth in the  
irreducible mass of each black hole, at a rate that should be
compatible with Eq.~(\ref{1.17}) if the holes are nonrotating. A
recent paper by Boyle {\it et al.}\ \cite{boyle-etal:07} suggests that  
the tidal heating of black holes might already have been seen in
numerical simulations --- witness their Figure 4.       

As an additional application of our work we mention another
connection with numerical relativity. If the simulations of black-hole
mergers are to describe realistic situations of astrophysical
interest, it is imperative that they proceed from data that correctly
describe the initial state of the system (the correct state of motion,
and the correct amount of initial radiation). The idea that
well-controlled initial-data sets could be constructed with the help
of post-Newtonian theory is an old one \cite{brady-etal:98,
  tichy-etal:03, blanchet:03, mora-will:02, mora-will:04, nissanke:06},
but its initial formulation did not account for the strong internal
gravity of each black hole, which cannot be approximated by a
post-Newtonian series. Alvi \cite{alvi:00, alvi:03} was the first to
remedy this situation by matching the post-Newtonian metric to
black-hole metrics, in the way that was reviewed in Secs.~I A and I
B. Alvi's work was improved upon by Yunes {\it et al.}\
\cite{yunes-etal:06} (see Sec.~I A), 
and our work contributes an additional improvement. While Alvi and
Yunes calculated the black-hole tidal moments for circular motion
only, and only to leading order in the post-Newtonian expansion, here
we calculate ${\cal E}_{ab}$ to first post-Newtonian order, and 
${\cal B}_{ab}$ to leading order, for more general situations. Our
results could be used to generate an improved version of the
Alvi-Yunes metric, which could then be used to construct improved 
initial-data sets for numerical relativity.  

\subsection{Organization of this paper}    

The technical portion of the paper begins in Sec.~II with an
introduction to the description of tidal environments in terms of STF
moments $\bar{\E}_{ab}$ and $\bar{\B}_{ab}$. In Sec.~III we present
the black-hole metric of Eqs.~(\ref{1.1}) and (\ref{1.2}). In Sec.~IV  
we introduce the post-Newtonian metric of
Eqs.~(\ref{1.3})--(\ref{1.5}), and discuss the decomposition of the 
gravitational potentials into a black-hole piece and an external
piece, as in Eq.~(\ref{1.6}). In Sec.~V we review the coordinate
transformation between the barycentric frame $(t,x^a)$ and the
black-hole frame $(\bar{t},\bar{x}^a)$, and we calculate how the 
post-Newtonian potentials change under this transformation. The end
result of this computation is a post-Newtonian metric expressed in the
same coordinates as the black-hole metric of Sec.~III. In Sec.~VI
we match the black-hole and post-Newtonian metrics and derive our
expression of Eq.~(\ref{1.7}) for the black hole's acceleration
vector, as well as Eqs.~(\ref{1.8}) and (\ref{1.9}) for the tidal
moments. In Sec.~VII we specialize our results to the specific case of
a two-body system. We first calculate the tidal moments for generic
orbital motion, and we next specialize these results to circular
motion; this gives rise to Eqs.~(\ref{1.10})--(\ref{1.16})
above. In Sec.~VIII we present an application of our results: We 
examine the intrinsic geometry of the event horizon of a
tidally-deformed black hole, in a suitable choice of gauge. And
finally, in Sec.~IX we apply our results to a calculation of
the tidal heating of a black hole by an external body on a circular
orbit; this is a gauge-independent effect. We first consider the case
of a nonrotating black hole and obtain Eq.~(\ref{1.17}). We next
consider the case of a rotating black hole; the result was not
displayed above, but it can be found in Eq.~(\ref{8.7}) below.  

Throughout the paper (except in Secs.~III B and VIII) we work in 
quasi-Lorentzian coordinates $x^\alpha = (ct,x^a)$ or $\bar{x}^\alpha
= (c\bar{t},\bar{x}^a)$, and we adopt a standard three-dimensional
notation when we deal with spatial components. For example, we use  
$\bm{v} = (v^x, v^y, v^z)$ to denote a Cartesian vector with
components $v^a$ in a flat, three-dimensional space. Indices on $v^a$ 
are manipulated with the Euclidean metric $\delta_{ab}$, and
$\epsilon_{abc}$ is the familiar permutation symbol. The Euclidean
norm of $\bm{v}$ is $|\bm{v}| := \sqrt{\delta_{ab} v^a v^b}$, and we
let $r := |\bm{x}|$ and $\bar{r} := |\bm{\bar{x}}|$. Because the paper
is devoted to a post-Newtonian treatment of tidal gravity, we find it
useful to use conventional units in which $G$ and $c$ are not set
equal to one (Sec.~III B is again an exception in this regard); as in
much of the literature on post-Newtonian theory, we use $c^{-2}$ as a
formal expansion parameter.      

\section{Tidal scales and tidal moments} 

As shown in Sec.~I B, the black hole's tidal environment is described
by the symmetric-tracefree (STF) tensors $\bar{\E}_{ab}(\bar{t})$ and 
$\bar{\B}_{ab}(\bar{t})$, and it is characterized by the length scale  
${\cal R}$. Our purpose in this section is to formally introduce these
quantities, and to set the stage toward the computation of the
black-hole metric in Sec.~III. The gravito-electric tidal moments
$\bar{\E}_{ab}$ can be introduced most simply in the context of
Newtonian gravity; we shall do this first. The gravito-magnetic
moments $\bar{\B}_{ab}$ do not exist in the Newtonian theory, but they
appear in a relativistic description of tidal fields. (The overbar, we
recall, indicates that the tidal moments are evaluated in a reference
frame that moves with the black hole.)   

In Newtonian gravity the total potential $U$ can be expressed as in
Eq.~(\ref{1.6}), with the first term $Gm/\bar{r}$ describing the black
hole (here $\bar{r}^2 := \delta_{ab} \bar{x}^a \bar{x}^b$), and the
second term $U_{\rm ext}$ describing the gravitational influence of
the external matter. Assuming that the external potential varies
slowly in the black-hole neighborhood $\scr B$, we express it as a
Taylor expansion in powers of $\bar{r}$: 
\begin{eqnarray*}
U_{\rm ext}(\bar{t}, \bar{x}^a) &=& U_{\rm ext}(\bar{t},0) 
+ \bar{g}_{a}(\bar{t}) \bar{x}^a 
+ \frac{1}{2} \bar{\E}_{ab}(\bar{t}) \bar{x}^a \bar{x}^b 
\nonumber \\ & & \mbox{} 
+ \frac{1}{6} \bar{\E}_{abc}(\bar{t}) \bar{x}^a \bar{x}^b \bar{x}^c  
+ O(\bar{r}^4/{\cal R}^4).
\end{eqnarray*}
Here $\bar{g}_a := \partial_{\bar{a}} U_{\rm ext}(\bar{t},0)$
is the gravitational force (per unit mass) acting on the black hole,
$\bar{\E}_{ab} := \partial_{\bar{a}\bar{b}} U_{\rm ext}(\bar{t},0)$ is
the quadrupole moment of the external potential, and $\bar{\E}_{abc}
:= \partial_{\bar{a}\bar{b}\bar{c}} U_{\rm ext}(\bar{t},0)$ is the
octupole moment. The first term in the expansion does not depend on
the spatial coordinates and plays no role in the gravitational
interaction of the black hole with the external distribution of
matter. The second term $\bar{g}_a \bar{x}^a$ also plays no role,
because we are working in a noninertial frame attached to the moving
black hole. What remains is the pure tidal potential, $U_{\rm tidal} 
= \frac{1}{2} \bar{\E}_{ab} \bar{x}^a \bar{x}^b + \frac{1}{6}
\bar{\E}_{abc} \bar{x}^a \bar{x}^b \bar{x}^c 
+ O(\bar{r}^4/{\cal R}^4)$, expressed in terms of the spatial
coordinates and the tidal moments (which depend on time). Notice that
the tidal moments are defined as fully symmetric tensors. And because
$U_{\rm ext}$ satisfies Laplace's equation in $\scr B$ (recall that
the black-hole neighborhood is assumed to be empty of matter), the
tidal moments are also tracefree. The quantities $\bar{\E}_{ab}$ and
$\bar{\E}_{abc}$ (and all higher-order moments) are therefore STF
tensors: $\bar{\E}_{ab} = \bar{\E}_\stf{ab}$ and $\bar{\E}_{abc} 
= \bar{\E}_\stf{abc}$. In addition, because $c^{-2} U_{\rm ext}$ is
dimensionless, $c^{-2} \bar{\E}_{ab}$ has a dimension of inverse
length squared, and the tidal length scale ${\cal R}$ is defined such
that the components of $c^{-2} \bar{\E}_{ab}$ are typically of order  
${\cal R}^{-2}$. In the preceding expression for $U_{\rm ext}$ we have
assumed that the components of $c^{-2} \bar{\E}_{abc}$ are of order 
${\cal R}^{-3}$, and that each additional term in the expansion comes
with an additional power of $\bar{r}/{\cal R}$.   

In general relativity the definition of the tidal moments
$\bar{\E}_{ab}$, $\bar{\E}_{abc}$, $\bar{\B}_{ab}$, $\bar{\B}_{abc}$
(and higher-order moments) requires more refinement. The relevant
tools were introduced by Thorne and Hartle \cite{thorne-hartle:85} and
Zhang \cite{zhang:86}. We consider a neighborhood of a geodesic world
line $\gamma$ in an arbitrary spacetime. In this neighborhood the
metric is assumed to satisfy the vacuum field equations. (In Sec.~III
a black hole will be placed on this geodesic, and our spacetime will 
become the ``external spacetime'' of Sec.~I A. For the time being,
however, the black hole is absent.) We install a normal coordinate
system $(\bar{t},\bar{x}^a)$ in the neighborhood. It possesses the 
following properties: (1) the spatial coordinates $\bar{x}^a$ vanish
on $\gamma$, and $\bar{t}$ is proper time on the geodesic; (2) the
metric takes on Minkowski values on $\gamma$; (3) all Christoffel
symbols vanish on $\gamma$; and (4) the coordinates are harmonic, in
the sense that the metric satisfies the conditions 
$\partial_\beta (\sqrt{-g} g^{\alpha\beta}) = 0$ everywhere in the
neighborhood. 

The metric near $\gamma$ admits an expansion in powers of
$\bar{r}$. By virtue of property (2) the zeroth-order terms are
constant, and by virtue of property (3) there are no terms at first
order. The terms at second and higher orders contain information about 
the curvature of spacetime near the geodesic, and it is those terms
that describe the tidal environment around $\gamma$. This environment
is characterized by the scaling quantities $\cal R$, $\cal L$, and
$\cal T$, with $\cal R$ denoting the radius of curvature on $\gamma$,   
$\cal L$ the scale of spatial inhomogeneity, and $\cal T$ the time
scale over which changes occur in the environment. The tidal
environment is described precisely by the tidal moments  
$\bar{\cal E}_{ab}(\bar{t}), \bar{\cal E}_{abc}(\bar{t}), \cdots$ and  
$\bar{\cal B}_{ab}(\bar{t}), \bar{\cal B}_{abc}(\bar{t}), \cdots$,
which are STF tensors that depend on $\bar{t}$ only. They are
related to components of the Riemann tensor and its derivatives
evaluated on $\gamma$ --- see Zhang's Eqs.~(1.3) for definitions
\cite{zhang:86}. The gravito-electric moments scale as 
\begin{eqnarray} 
c^{-2} \bar{\cal E}_{ab} &\sim& {\cal R}^{-2},
\nonumber \\  
c^{-2} \bar{\cal E}_{abc} &\sim& {\cal R}^{-2} {\cal L}^{-1},
\label{2.1}
\\ 
c^{-2} \dot{\bar{\cal E}}_{ab} &\sim& {\cal R}^{-2} {\cal T}^{-1},  
\nonumber 
\end{eqnarray} 
in which an overdot indicates differentiation with respect to
$\bar{t}$. It is the relations of Eq.~(\ref{2.1}) that define the
scales $\cal R$, $\cal L$, and $\cal T$. The gravito-magnetic moments
scale as 
\begin{eqnarray} 
c^{-3} \bar{\cal B}_{ab} &\sim& (v/c) {\cal R}^{-2},
\nonumber \\  
c^{-3} \bar{\cal B}_{abc} &\sim& (v/c) {\cal R}^{-2} {\cal L}^{-1}, 
\\
c^{-3} \dot{\bar{\cal B}}_{ab} &\sim& (v/c) {\cal R}^{-2} 
{\cal T}^{-1}, 
\nonumber 
\label{2.2}
\end{eqnarray} 
in which $v \sim {\cal L}/{\cal T}$ is a velocity scale. In a
slow-motion context we have that $v/c \ll 1$, and the gravito-magnetic
moments are smaller than the gravito-electric moments by a small
factor of order $v/c$.  

To illustrate the meaning of these tidal scales we return to the
example presented in Sec.~I A, in which the tidal environment is
provided by an external body of mass $m'$. We suppose that the
geodesic is at a distance $b$ from the external body. In this
situation $v/c \sim \sqrt{M'/b}$, where $M' := Gm'/c^2$ is the
characteristic gravitational radius of the external body. The tidal
scales are then given by   
\[
{\cal R} \sim \sqrt{b^3/M'}, \qquad 
{\cal L} \sim b, \qquad 
{\cal T} \sim \sqrt{b^3/Gm'}.
\]
We notice that ${\cal L} \sim (v/c) {\cal R}$ and ${\cal T} \sim 
{\cal R}/c$; in a slow-motion situation we have that ${\cal L} \ll  
{\cal R}$.  

The metric near $\gamma$ takes the form derived by Zhang
\cite{zhang:86}  
\begin{eqnarray} 
g_{\bar{0}\bar{0}} &=& -1 
- \frac{1}{c^2} \bar{\cal E}_{ab} \bar{x}^a \bar{x}^b   
- \frac{1}{3c^2} \bar{\cal E}_{abc} \bar{x}^a \bar{x}^b \bar{x}^c 
+ \cdots, 
\label{2.3} 
\\ 
g_{\bar{0}\bar{a}} &=& 
\frac{2}{3c^3} \epsilon_{abp} \bar{\cal B}^p_{\ c}  
  \bar{x}^b \bar{x}^c 
+ \frac{1}{3c^3} \epsilon_{abp} \bar{\cal B}^p_{\ cd}
  \bar{x}^b \bar{x}^c \bar{x}^d 
\nonumber \\ & & \mbox{}
- \frac{10}{21c^3} \Bigl( \bar{x}_a \dot{\bar{\cal E}}_{bc} 
  \bar{x}^b \bar{x}^c 
- \frac{2}{5} \bar{r}^2 \dot{\bar{\cal E}}_{ab} \bar{x}^b \Bigr) 
+ \cdots, 
\label{2.4} 
\\ 
g_{\bar{a}\bar{b}} &=& \delta_{ab} \biggl( 1 
+ \frac{1}{c^2} \bar{\cal E}_{cd} \bar{x}^c \bar{x}^d  
+ \frac{1}{3c^2} \bar{\cal E}_{cde} \bar{x}^c \bar{x}^d \bar{x}^e 
\biggr) 
\nonumber \\ & & \mbox{}
+ \frac{5}{21c^4} \Bigl( \bar{x}_{(a} \epsilon_{b)cp} 
  \dot{\bar{\cal B}}^p_{\ d} \bar{x}^c \bar{x}^d 
- \frac{1}{5} \bar{r}^2 \epsilon_{cp(a} \dot{\bar{\cal B}}^p_{\ b)}
\bar{x}^c \Bigr) 
\nonumber \\ & & \mbox{}
+ \cdots,
\label{2.5} 
\end{eqnarray} 
where $\epsilon_{abc}$ is the permutation symbol. The neglected terms
involve higher powers of $\bar{r}$, and higher-order tidal moments.  

The first tidal term on the right-hand side of Eq.~(\ref{2.3}) is of
order $(\bar{r}/{\cal R})^2$, and the second term is smaller than this
by a factor of order $\bar{r}/{\cal L}$; the neglected terms are
smaller still, by additional factors of order $\bar{r}/{\cal L}$. In
Eq.~(\ref{2.4}) the first term is of order 
$(v/c) (\bar{r}/{\cal R})^2$, and the second term is smaller than this
by a factor of order $\bar{r}/{\cal L}$; taking into account the
scalings ${\cal L} \sim (v/c) {\cal R}$ and 
${\cal T} \sim {\cal R}/c$, the same is true of the third and fourth
terms. In Eq.~(\ref{2.5}) the first tidal term is of order 
$(\bar{r}/{\cal R})^2$, and the second term is smaller by a factor of
order $\bar{r}/{\cal L}$; the third and fourth terms are smaller than
this by another factor of order $\bar{r}/{\cal L}$, and they come also
with an additional factor of order $(v/c)^2$.    

These considerations lead us to the following conclusion: If we
restrict the neighborhood of $\gamma$ to be such that $\bar{r}$ is
everywhere much smaller than ${\cal L}$, then Zhang's metric can be
simplified to  
\begin{eqnarray} 
g_{\bar{0}\bar{0}} &=& -1 
- \frac{1}{c^2} \bar{\cal E}_{ab} \bar{x}^a \bar{x}^b  
+ O\biggl( \frac{\bar{r}^3}{{\cal R}^2 {\cal L}} \biggr), 
\label{2.6} \\ 
g_{\bar{0}\bar{a}} &=& 
\frac{2}{3c^3} \epsilon_{abp} \bar{\cal B}^p_{\ c} \bar{x}^b \bar{x}^c 
+ O\biggl( \frac{v}{c} \frac{\bar{r}^3}{{\cal R}^2 {\cal L}} \biggr), 
\label{2.7} \\ 
g_{\bar{a}\bar{b}} &=& \delta_{ab} \biggl( 1 
+ \frac{1}{c^2} \bar{\cal E}_{cd} \bar{x}^c \bar{x}^d \biggr) 
+ O\biggl( \frac{\bar{r}^3}{{\cal R}^2 {\cal L}} \biggr).
\label{2.8} 
\end{eqnarray} 
This simplified form, which involves the lowest-order tidal moments
only, and which neglects their time derivatives, shall be sufficient
for our purposes below. 

\section{Tidally deformed black hole} 

\subsection{Metric of a deformed black hole} 

The metric of Eqs.~(\ref{2.6})--(\ref{2.8}) describes the tidal
environment around a geodesic $\gamma$ in an arbitrary spacetime, with
the only restriction that the geodesic's neighborhood must be empty of 
matter. We now place a nonrotating black hole of mass $m$ on this
geodesic, and modify the metric to account for its gravitational
effects. As we shall show below (in Sec.~III B), the metric of the
tidally deformed black hole is given by  
\begin{widetext}
\begin{eqnarray} 
g_{\bar{0}\bar{0}} &=& -\frac{1-M/\bar{r}}{1+M/\bar{r}} 
- \bar{r}^2(1-M/\bar{r})^2 \E^{\sf q} 
+ O\biggl( \frac{\bar{r}^3}{{\cal R}^2 {\cal L}} \biggr),  
\label{3.1} \\  
g_{\bar{0}\bar{a}} &=& \frac{2}{3} \bar{r}^2 
(1-M/\bar{r})(1+M/\bar{r})^2 \B^{\sf q}_a 
+ O\biggl( \frac{v}{c} \frac{\bar{r}^3}{{\cal R}^2 {\cal L}} \biggr),  
\label{3.2} \\ 
g_{\bar{a}\bar{b}} &=& \frac{1+M/\bar{r}}{1-M/\bar{r}} \Omega_a \Omega_b  
+ (1+M/\bar{r})^2 \gamma_{ab} 
- \bar{r}^2(1 + M/\bar{r})^2 \E^{\sf q} \Omega_a \Omega_b 
\nonumber \\ & & \mbox{}
- M\bar{r}(1+M/\bar{r})^2 (1+M^2/3\bar{r}^2) 
  (\Omega_a \E^{\sf q}_b + \E^{\sf q}_a \Omega_b) 
- \bar{r}^2 (1-M/\bar{r})^2(1+M/\bar{r})^3 
  \gamma_{ab} \E^{\sf q} 
\nonumber \\ & & \mbox{}
- M \bar{r} (1+M/\bar{r})^2 (1-M^2/3\bar{r}^2) \E^{\sf q}_{ab}
+ O\biggl( \frac{\bar{r}^3}{{\cal R}^2 {\cal L}} \biggr)
\label{3.3}   
\end{eqnarray} 
\end{widetext}
in the hole's neighborhood $\scr B$, which is formally defined by
$\bar{r} < \bar{r}_{\rm max}$ with $\bar{r}_{\rm max} \ll {\cal L}$
(this is a refinement of the definition presented in Sec.~I B). Here
$M := Gm/c^2$ is the black hole's gravitational radius, 
$\Omega^a := \bar{x}^a/\bar{r}$ is a unit radial vector, and 
$\gamma_{ab} := \delta_{ab} - \Omega_a \Omega_b$. The tidal potentials
are given by   
\begin{eqnarray} 
\E^{\sf q} &=& \frac{1}{c^2} \bar{\E}_{cd} \Omega^c \Omega^d, 
\label{3.4} \\ 
{\cal E}^{\sf q}_a &=& \frac{1}{c^2} \gamma_a^{\ c} \bar{\E}_{cd}
\Omega^d,  
\label{3.5} \\ 
\E^{\sf q}_{ab} &=& \frac{1}{c^2} \Bigl( 2\gamma_a^{\ c} \gamma_b^{\ d} 
\bar{\E}_{cd} + \gamma_{ab} \E^{\sf q} \Bigr),
\label{3.6} \\ 
\B^{\sf q}_a &=& \frac{1}{c^3} \epsilon_{apq} \Omega^p  
\bar{\B}^q_{\ c} \Omega^c; 
\label{3.7} 
\end{eqnarray}
the label $\sf q$ stands for ``quadrupole.'' The metric of
Eqs.~(\ref{3.1})--(\ref{3.3}) is presented in harmonic coordinates,
and it is an approximate solution to the vacuum field equations
linearized about the Schwarzschild metric. Indeed, setting
$\bar{\E}_{ab} = \bar{\B}_{ab} = 0$ in Eqs.~(\ref{3.1})--(\ref{3.3})
returns the Schwarzschild metric in harmonic coordinates, and the
tidal potentials represent a pure-quadrupole metric perturbation. It
is easy to see that when $\bar{r}$ is much larger than $M$ (but still
much smaller than ${\cal L}$), Eqs.~(\ref{3.1})--(\ref{3.3}) reduce to 
Eqs.~(\ref{2.6})--(\ref{2.8}). This shows that Zhang's metric provides
the appropriate asymptotic conditions for the metric of a tidally
deformed black hole; these replace the asymptotically-flat conditions
that would be appropriate for an isolated black hole. Like the metric
of Eqs.~(\ref{2.6})--(\ref{2.8}), the black-hole metric neglects tidal
terms that are smaller than the dominant ones by additional factors of
order $\bar{r}/{\cal L} \ll 1$; within this approximation the metric
is accurate to all orders in $M/\bar{r}$.    

The method by which the metric of Eqs.~(\ref{3.1})--(\ref{3.3}) is 
obtained is explained in the next subsection. The reader not
interested in those details can immediately proceed to Sec.~IV, in
which we present the post-Newtonian metric to which the black-hole
metric will be matched. Before we proceed, however, it is useful to
note that the components of the black-hole metric that are required
for matching are  
\begin{eqnarray} 
g_{\bar{0}\bar{0}} &=& -\frac{1-M/\bar{r}}{1+M/\bar{r}} 
- \frac{1}{c^2} (1-M/\bar{r})^2 \bar{\E}_{ab}(\bar{t})  
  \bar{x}^a \bar{x}^b 
\nonumber \\ & & \mbox{}
+ O\biggl( \frac{\bar{r}^3}{{\cal R}^2 {\cal L}} \biggr)  
\label{3.8}
\end{eqnarray} 
and 
\begin{eqnarray} 
g_{\bar{0}\bar{a}} &=& \frac{2}{3c^3} (1-M/\bar{r})(1+M/\bar{r})^2   
  \epsilon_{abp} \bar{\B}^p_{\ c}(\bar{t}) 
  \bar{x}^b \bar{x}^c 
\nonumber \\ & & \mbox{}
+ O\biggl( \frac{v}{c} \frac{\bar{r}^3}{{\cal R}^2 {\cal L}} \biggr). 
\label{3.9} 
\end{eqnarray} 
We recall that $M = Gm/c^2$ is the black hole's gravitational 
radius. 

\subsection{Derivation} 

We begin this subsection with a warning on a change of notation; this
change applies to this subsection only. To simplify the notation we
shall refrain from displaying the overbar, in spite of the fact that
we continue to work in a reference frame that moves with the black
hole. In addition, we shall use relativist's units, in which $c$ and
$G$ are both set equal to unity. And finally, in this section we use 
$g_{\alpha\beta}$ to denote the Schwarzschild metric (as opposed to
the perturbed metric displayed in Sec.~III A), and we let
$h_{\alpha\beta}$ denote the tidal perturbation; the perturbed metric
is therefore $g_{\alpha\beta} + h_{\alpha\beta}$.   

To obtain Eqs.~(\ref{3.1})--(\ref{3.7}) we rely on
Ref.~\cite{poisson:05}, in which the metric of a tidally deformed,
nonrotating black hole is presented in light-cone coordinates
$(v,\rho,\theta,\phi)$. The advanced-time coordinate $v$ is constant
on past light cones that converge toward $\rho = 0$, $\rho$ is the
usual Schwarzschild radial coordinate that measures the area of closed
surfaces of constant $(v,\rho)$, and $\theta^A = (\theta,\phi)$ are
angular coordinates on these surfaces. In the light-cone coordinates
the Schwarzschild metric takes the form 
\begin{equation} 
g_{\alpha\beta} dx^\alpha dx^\beta = -f\, dv^2 
+ 2\, dvd\rho + \rho^2 \Omega_{AB}\, d\theta^A d\theta^B, 
\label{3.10} 
\end{equation} 
with $f := 1-2M/\rho$ and $\Omega_{AB}\, d\theta^A d\theta^B 
:= d\theta^2 + \sin^2\theta\, d\phi^2$ denoting the metric on the unit
two-sphere. The nonvanishing components of the tidal perturbation are 
\begin{eqnarray} 
h^{\rm light}_{vv} &=& -\rho^2 f^2 \E^{\sf q}, 
\label{3.11} \\ 
h^{\rm light}_{vA} &=& -\frac{2}{3} \rho^3 f \big( \E^{\sf q}_A 
- \B^{\sf q}_A \bigr), 
\label{3.12} \\ 
h^{\rm light}_{AB} &=& 
-\frac{1}{3} \rho^4 \bigl( 1 - 2 M^2/\rho^2 \bigr) \E^{\sf q}_{AB} 
\nonumber \\ & & \mbox{}
+ \frac{1}{3} \rho^4 \bigl( 1 - 6 M^2/\rho^2 \bigr) \B^{\sf q}_{AB}, 
\label{3.13}
\end{eqnarray} 
in which the label ``light'' indicates that the perturbation is
presented in the light-cone gauge. The tidal potentials are given in
terms of the tidal moments $\E_{ab}(v)$ and $\B_{ab}(v)$ by the
relations    
\begin{eqnarray} 
\E^{\sf q} &=& \E_{ab} \Omega^a \Omega^b, 
\label{3.14} \\ 
\E^{\sf q}_A &=& \Omega^a_A \E_{ab} \Omega^b, 
\label{3.15} \\ 
\E^{\sf q}_{AB} &=& 2\Omega^a_A \Omega^b_B \E_{ab} 
+ \Omega_{AB} \E^{\sf q}, 
\label{3.16} \\ 
\B^{\sf q}_A &=& \Omega^a_A \epsilon_{apq} \Omega^p \B^q_{\ b}
\Omega^b, 
\label{3.17} \\ 
\B^{\sf q}_{AB} &=& \Omega^a_A \epsilon_{apq} \Omega^p \B^q_{\ b}
\Omega^b_B + \Omega^b_B \epsilon_{bpq} \Omega^p \B^q_{\ a} 
\Omega^a_A, \quad
\label{3.18} 
\end{eqnarray} 
where $\Omega^a := (\sin\theta\cos\phi,\sin\theta\sin\phi,\cos\theta)$
and $\Omega^a_A := \partial \Omega^a/\partial \theta^A$. The perturbed 
metric is $g_{\alpha\beta} + h_{\alpha\beta}$, and it is
straightforward to show that this metric satisfies the vacuum field
equations linearized about the Schwarzschild solution. The metric is
accurate up to terms of order $(r^3/{\cal R}^2 {\cal L})$ that come
from higher-order tidal moments, and terms of order      
$(r^3/{\cal R}^2 {\cal T})$ that come from the time derivative
of the quadrupole moments. For the purposes of substitution into the
field equations, $\E_{ab}$ and $\B_{ab}$ can be considered to be
time-independent.  

The black-hole metric is expressed in the light-cone coordinates
$(v,\rho,\theta^A)$, and we wish to perform a transformation to
harmonic coordinates $(t,x^a)$. We accomplish this in two
steps. First, we perform a {\it gauge transformation} to
change the perturbation $h_{\alpha\beta}$ from its current
light-cone gauge to a harmonic gauge. Second, we perform a
{\it background coordinate transformation} from the (background)
light-cone coordinates to the (background) harmonic coordinates. This
strategy, and its implementation detailed below, was suggested to
us by Steve Detweiler (personal communication). 

Let a set of four scalar fields be defined by
\begin{eqnarray} 
T &:=& v - \rho - 2M\log(\rho/2M - 1),
\label{3.19} \\ 
X &:=& (\rho-M) \sin\theta\cos\phi, 
\label{3.20} \\  
Y &:=& (\rho-M) \sin\theta\sin\phi,
\label{3.21} \\  
Z &:=& (\rho-M) \cos\theta, 
\label{3.22}  
\end{eqnarray} 
and let us collectively denote the members of this set by
$X^{(\mu)}$. It is straightforward to show that each one of the scalar
fields $X^{(\mu)}$ satisfies the wave equation 
\begin{equation} 
\Box X^{(\mu)} := g^{\alpha\beta} \nabla_\alpha \nabla_\beta
X^{(\mu)} = \frac{1}{\sqrt{-g}} \partial_\beta \bigl(
g^{\alpha\beta} \partial_\alpha X^{(\mu)} \bigr) = 0
\label{3.23}
\end{equation} 
in the Schwarzschild spacetime. Here $\nabla_\alpha$ is the
covariant-derivative operator compatible with the Schwarzschild metric 
$g_{\alpha\beta}$, and $g$ is the metric determinant. The statement of
Eq.~(\ref{3.23}) is coordinate independent. When, however, we choose
$(t=T, x=X, y=Y, z=Z)$ as coordinates, then Eq.~(\ref{3.23}) becomes 
$\partial_\beta (\sqrt{-g} g^{\mu\beta}) = 0$, the familiar
statement of the harmonic coordinate condition. Equation (\ref{3.23})
therefore provides a coordinate-invariant way of stating that the
scalar fields $X^{(\mu)}$ form a set of harmonic coordinates for the
Schwarzschild spacetime.  

We now demand that $X^{(\mu)}$ be harmonic coordinates for 
the perturbed spacetime, {\it in addition to} being harmonic
coordinates for the Schwarzschild spacetime. To achieve this we
rewrite Eq.~(\ref{3.23}) in terms of the full metric $g_{\alpha\beta}
+ h_{\alpha\beta}$ and its associated covariant-derivative
operator. We write $g^{\alpha\beta} - h^{\alpha\beta}$ for the inverse 
metric and $\sqrt{-g} (1 + \frac{1}{2} h)$ for the metric determinant,
and we manipulate the indices with the background metric; for example
$h^{\alpha\beta} = g^{\alpha\gamma} g^{\beta\delta} h_{\gamma\delta}$
and $h = g^{\alpha\beta} h_{\alpha\beta}$. After some straightforward
manipulations we find that in addition to Eq.~(\ref{3.23}), the scalar
fields must also satisfy the set of equations 
\begin{equation} 
\nabla_\alpha \bigl( 
\psi^{\alpha\beta} \nabla_\beta X^{(\mu)} \bigr) 
= 0, 
\label{3.24}
\end{equation}
where $\psi_{\alpha\beta} = h_{\alpha\beta} - \frac{1}{2}
g_{\alpha\beta} h$ is the ``trace-reversed'' metric perturbation. This
equation can be interpreted as a gauge condition on $h_{\alpha\beta}$:
The perturbation will be in the {\it harmonic gauge} if it satisfies 
the conditions of Eq.~(\ref{3.24}). This equation is coordinate
invariant, and the harmonic-gauge condition can be imposed even when 
the coordinates $x^\alpha$ do not coincide with the scalar fields
$X^{(\mu)}$. 

In Eq.~(\ref{3.11})--(\ref{3.13}) the metric perturbation is presented
in the light-cone gauge. A transformation to the harmonic gauge will
be generated by the vector field $\xi_\alpha$, so that 
\begin{equation} 
h_{\alpha\beta}^{\rm harm} = h_{\alpha\beta}^{\rm light} 
- \nabla_\alpha \xi_\beta - \nabla_\beta \xi_\alpha. 
\label{3.25}
\end{equation} 
It is straightforward to show that if $h_{\alpha\beta}^{\rm harm}$ is
to satisfy Eq.~(\ref{3.24}), then the vector field must satisfy the
four equations 
\begin{eqnarray}
\nabla_\alpha \bigl( \psi_{\rm light}^{\alpha\beta}
\nabla_\beta X^{(\mu)} \bigr) &=& 
\bigl( \Box \xi^\alpha \bigr) \nabla_\alpha X^{(\mu)} 
\nonumber \\ & & \mbox{} 
+ 2 \bigl( \nabla^\alpha \xi^\beta \bigr) \nabla_\alpha \nabla_\beta
X^{(\mu)}; \qquad 
\label{3.26}
\end{eqnarray} 
in our case the perturbation is traceless in the light-cone gauge, so
$\psi^{\rm light}_{\alpha\beta} = h^{\rm light}_{\alpha\beta}$. It may
be verified that the vector  
\begin{eqnarray} 
\xi_v &=& -\frac{1}{3} \rho^3 f \E^{\sf q}, 
\label{3.27} \\
\xi_\rho &=& \frac{1}{3} \rho^3 \E^{\sf q}, 
\label{3.28} \\ 
\xi_A &=& -\frac{ \rho^5 f^2 }{3(\rho-M)} \E^{\sf q}_A 
+ \frac{1}{3} \rho^2 (\rho^2 - 6M^2) \B^{\sf q}_A \quad
\label{3.29} 
\end{eqnarray}
is a solution to Eqs.~(\ref{3.26}). In this computation the tidal
moments $\E_{ab}$ and $\B_{ab}$ can be considered to be
time-independent, because the time derivative of $\xi_\alpha$ is
smaller than its spatial derivatives by a small factor of order  
$\rho/{\cal T}$. 

Substitution of Eqs.~(\ref{3.11})--(\ref{3.13}) and
Eqs.~(\ref{3.27})--(\ref{3.29}) into Eq.~(\ref{3.25}) returns the
tidal perturbation in the desired harmonic gauge. We obtain 
\begin{eqnarray} 
h^{\rm harm}_{vv} &=& -\rho^2 f^2 \E^{\sf q}, 
\label{3.30} \\
h^{\rm harm}_{v\rho} &=& \rho^2 f \E^{\sf q}, 
\label{3.31} \\
h^{\rm harm}_{\rho\rho} &=& -2\rho^2 \E^{\sf q}, 
\label{3.32} \\
h^{\rm harm}_{vA} &=& \frac{2}{3} \rho^3 f \B^{\sf q}_A, 
\label{3.33} \\
h^{\rm harm}_{\rho A} &=& -\frac{M\rho^2}{3(\rho-M)^2}  
( 3\rho^2 - 6M\rho + 4M^2 ) \E^{\sf q}_A 
\nonumber \\ & & \mbox{} 
- \frac{2}{3} \rho^3 \B^{\sf q}_A, 
\label{3.34} \\ 
h^{\rm harm}_{AB} &=& 
-\frac{\rho^5 f^2}{\rho-M} \Omega_{AB} \E^{\sf q} 
\nonumber \\ & & \mbox{} 
- \frac{M\rho^2}{3(\rho-M)}  
( 3\rho^2 - 6M\rho + 2M^2 ) \E^{\sf q}_{AB}. \qquad 
\label{3.35}
\end{eqnarray} 
The full metric is next obtained by adding $h_{\alpha\beta}$ to
$g_{\alpha\beta}$ as given by Eq.~(\ref{3.10}). 

The tidal perturbation is now correctly expressed in the harmonic
gauge, but the metric is still written in terms of the original
coordinates $(v,\rho,\theta,\phi)$. Our final step is therefore to  
perform a coordinate transformation from these coordinates to the
harmonic coordinates $(t=T, x=X, y=Y, z=Z)$. We carry this out in two
stages. First, we effect a transformation from $(v,\rho)$
to $(t,r)$, leaving the angular coordinates alone; here $t = v - \rho
- 2M\log(\rho/2M - 1)$ is harmonic time and $r = \rho - M$ is the 
harmonic radial coordinate. This coordinate transformation brings the 
Schwarzschild metric to the new form 
\begin{eqnarray} 
g_{\alpha\beta}\, dx^\alpha dx^\beta &=& 
-\frac{1-M/r}{1+M/r}\, dt^2 
+ \frac{1+M/r}{1-M/r}\, dr^2 
\nonumber \\ & & \mbox{} 
+ (r+M)^2\, \Omega_{AB}\, d\theta^A d\theta^B, 
\label{3.36}
\end{eqnarray}
and the tidal perturbation becomes
\begin{eqnarray} 
h^{\rm light}_{tt} &=& -r^2(1-M/r)^2 \E^{\sf q}, 
\label{3.37} \\ 
h^{\rm light}_{rr} &=& -r^2(1+M/r)^2 \E^{\sf q}, 
\label{3.38} \\ 
h^{\rm light}_{tA} &=& \frac{2}{3} r^3 (1-M/r)(1+M/r)^2 \B^{\sf q}_A,
\label{3.39} \\ 
h^{\rm light}_{rA} &=& - M r^2 (1+M/r)^2 (1+M^2/3r^2) \E^{\sf q}_A,
\label{3.40} \\ 
h^{\rm light}_{AB} &=& -r^4 (1-M/r)^2 (1+M/r)^3 \Omega_{AB} \E^{\sf q} 
\nonumber \\ & & \mbox{} 
- M r^3 (1+M/r)^2 (1-M^2/3r^2) \E^{\sf q}_{AB}. \qquad
\label{3.41}
\end{eqnarray} 
In the second stage we go from the spherical coordinates
$(r,\theta^A)$ to the associated Cartesian coordinates $x^a =
r\Omega^a$, that is, $x = r\sin\theta\cos\phi$,
$y=r\sin\theta\sin\phi$, and $z = r\cos\theta$. The transformation
matrix is $\partial r/\partial x^a = \Omega_a$ and 
$\partial \theta^A / \partial x^a = r^{-1} \Omega^A_a$, in which 
$\Omega_a := \delta_{ab} \Omega^b$ and $\Omega^A_a 
:= \Omega^{AB} \delta_{ab} \Omega^b_B$. The transformation  
of $g_{\alpha\beta} + h_{\alpha\beta}$ under this change of
coordinates is easy to carry out, and the end result is the
metric of Eqs.~(\ref{3.1})--(\ref{3.3}). 

\section{Post-Newtonian metric} 

In this section we take our attention away from the black-hole
neighborhood ${\scr B}$, and we take the global view of the
post-Newtonian domain ${\scr D}$, which was introduced in 
Sec.~I B. Let us recall that ${\scr D}$ is spatially limited by a
sphere of radius $r_{\rm near}$ centered on the post-Newtonian
barycenter, and that this sphere marks the boundary of the near
zone. The domain excludes a sphere of radius $\bar{r}_{\rm min}$
centered on the black hole, inside which the hole's gravity is too
strong to be adequately described by post-Newtonian theory. We recall
also that there exists an overlap ${\scr O}$ between the black-hole
neighborhood ${\scr B}$ and the post-Newtonian domain $\scr D$. This
is described by $\bar{r}_{\rm min} < \bar{r} < \bar{r}_{\rm max}$, in
which $\bar{r}_{\rm max} \ll {\cal L}$ is the boundary of 
${\scr B}$. We assume that there is no matter in ${\scr O}$.  

The metric in ${\scr D}$ is expressed in the post-Newtonian form  
\begin{eqnarray} 
g_{00} &=& -1 + \frac{2}{c^2} U + \frac{2}{c^4} (\Psi - U^2) 
+ O(c^{-6}), 
\label{4.1} \\ 
g_{0a} &=& -\frac{4}{c^3} U_a + O(c^{-5}), 
\label{4.2} \\
g_{ab} &=& \biggl( 1 + \frac{2}{c^2} U \biggr) \delta_{ab}  
+ O(c^{-4}), 
\label{4.3}
\end{eqnarray} 
in which $U$ is a Newtonian potential, $U_a$ a vector potential, and
$\Psi$ a post-Newtonian potential. The metric is presented in harmonic 
coordinates $x^\alpha = (ct, x^a)$. These coordinates are centered on
the post-Newtonian barycenter, which defines the origin of an inertial
reference frame; we recall that they are distinct from the harmonic
coordinates $\bar{x}^\alpha = (c\bar{t}, \bar{x}^a)$ used in 
${\scr B}$. 

We assume that there is a distribution of matter within $\scr D$, and
we make only two assumptions regarding its nature. First, we assume
that its gravity is sufficiently weak to be adequately described by
the post-Newtonian metric of Eqs.~(\ref{4.1})--(\ref{4.3}). Second, we
assume that there is no matter in $\scr O$, so that the
black hole moves in an empty region of spacetime. Otherwise the
distribution of matter within $\scr D$ is completely arbitrary. It
could be a continuous fluid, so as to model an accretion disk or a
galactic core; or it could correspond to a collection of $N-1$ bodies
with weak self-gravity, making the black hole a member of an $N$-body
system; or else the domain $\scr D$ could exclude a number $N-1$ of
small regions that would each contain a condensed body such as a
neutron star or a black hole. Our considerations allow for this degree 
of generality.  

We wish to solve the Einstein field equations in the vacuum domain
${\scr O}$. To write them down it is convenient to express the
post-Newtonian potential as  
\begin{equation} 
\Psi = \psi + \frac{1}{2} \frac{\partial^2 X}{\partial t^2}, 
\label{4.4}
\end{equation} 
in terms of two new potentials $\psi$ and $X$. The degeneracy is
broken by the field equations, which are given by (see, for example,
Ref.~\cite{racine-flanagan:05}) 
\begin{eqnarray} 
\nabla^2 U &=& 0, 
\label{4.5} \\ 
\nabla^2 U^a &=& 0, 
\label{4.6} \\ 
\nabla^2 \psi &=& 0, 
\label{4.7} \\ 
\nabla^2 X &=& 2U,  
\label{4.8} 
\end{eqnarray} 
where $\nabla^2$ is the usual Laplacian operator of three-dimensional
flat space. The potential $X$ is commonly referred to as the
post-Newtonian superpotential. The vacuum field equations are
augmented by the harmonic coordinate condition 
\begin{equation} 
\partial_t U + \partial_a U^a = 0, 
\label{4.9}
\end{equation} 
which takes the form of a gauge condition on the gravitational 
potentials. 

Each one of the field equations is linear, and a solution representing
a black hole and an external distribution of matter can be obtained by
linear superposition. We treat the black hole as a post-Newtonian
monopole of mass $m$ moving on a trajectory described by the position
vector $\bm{x} = \bm{z}(t)$; we let $\bm{v} := d\bm{z}/dt$ be the
black hole's velocity vector, and $\bm{a} := d\bm{v}/dt$ is the
acceleration vector. In this treatment the solutions to the field
equations are expressed as 
\begin{eqnarray} 
U(t,\bm{x}) &=& \frac{Gm}{|\bm{x}-\bm{z}|} 
+ U_{\rm ext}(t,\bm{x}),  
\label{4.10} \\ 
U^a(t,\bm{x}) &=& \frac{Gmv^a}{|\bm{x}-\bm{z}|} 
+ U^a_{\rm ext}(t,\bm{x}),
\label{4.11} \\  
\psi(t,\bm{x}) &=& \frac{Gm \mu}{|\bm{x}-\bm{z}|} 
+ \psi_{\rm ext}(t,\bm{x}),
\label{4.12} \\  
X(t,\bm{x}) &=& Gm |\bm{x}-\bm{z}| + X_{\rm ext}(t,\bm{x}), 
\label{4.13}
\end{eqnarray} 
where $|\bm{x} - \bm{z}(t)|$ is the Euclidean distance between
the field point $\bm{x}$ and the black hole. 

The first term on the right-hand side of each equation represents the
black hole, and we see that indeed, each black-hole potential has a
monopole structure. The fact that the vector $\bm{v}$ appears in the
vector potential $U^a$ is a consequence of the gauge condition of
Eq.~(\ref{4.9}). The quantity $\mu(t)$ that appears in $\psi$ is a
post-Newtonian correction to the mass parameter $m$; this cannot be
determined directly from the vacuum field equations.\footnote{If the
black hole were treated as a point particle, solving the field
equations in the presence of a distributional energy-momentum tensor
would reveal that $\mu = \frac{3}{2} v^2 - U_{\rm ext}(t,\bm{x} =
\bm{z})$, where $v^2 = |\bm{v}|^2$. This is indeed the correct
expression, but in our approach the determination of $\mu$ will come
at a later stage, from a careful matching of the post-Newtonian metric
with the black-hole metric of Sec.~III. For the time being $\mu(t)$
will remain undetermined.}

The second terms on the right-hand sides of
Eqs.~(\ref{4.10})--(\ref{4.13}) are the potentials created by the
matter distribution external to $\scr O$. In $\scr O$ they separately 
satisfy the vacuum field equations of Eqs.~(\ref{4.5})--(\ref{4.8}),
and they separately satisfy the harmonic-gauge condition of
Eq.~(\ref{4.9}),  
\begin{equation} 
\partial_t U_{\rm ext} + \partial_a U^a_{\rm ext} = 0. 
\label{4.14}
\end{equation}
We assume that the external potentials are smooth functions of the
coordinates in a neighborhood of $\bm{x} = \bm{z}(t)$.  

From Eq.~(\ref{4.4}) we find that the post-Newtonian potential $\Psi$
is given by  
\begin{equation} 
\Psi = -\frac{Gm}{2 s^3} (\bm{v} \cdot \bm{s})^2 
+ \frac{Gm}{s} \Bigl( \mu + \frac{1}{2} v^2 \Bigr) 
- \frac{Gm}{2s} \bm{a} \cdot \bm{s} + \Psi_{\rm ext}, 
\label{4.15}
\end{equation} 
where $\bm{s} := \bm{x} - \bm{z}(t)$, $s := |\bm{s}|$, and 
$\Psi_{\rm ext} = \psi_{\rm ext} 
+ \frac{1}{2} \partial_t^2 X_{\rm ext}$.  

The foregoing equations provide necessary and sufficient information
regarding the post-Newtonian environment of the black hole. A
remarkable and important aspect of our discussion is that the black
hole is treated as a post-Newtonian monopole. This feature requires a 
justification. At the level of Eqs.~(\ref{4.10})--(\ref{4.13}) the
monopole nature of the black hole is introduced as an assumption. As
we proceed with the matching of the post-Newtonian metric with the
black-hole metric of Sec.~III, however, we shall see that our
assumption is the only one that is consistent with the given structure
of the black-hole metric. If we had instead given an arbitrary
multipole structure to our black hole, the matching procedure would
eventually force us to set all higher multipole moments to zero; the
only surviving moment is the black-hole mass $m$. As we shall see, 
therefore, {\it the statement that the black
hole is a post-Newtonian monopole is a strict consequence of the field
equations of general relativity; no other multipole structure is
possible.} The monopole structure is not assumption introduced for its
simplicity; it is a direct outcome of the matching procedure.  

\section{Transformation from barycenter frame to black-hole frame}   

The post-Newtonian metric of Eqs.~(\ref{4.1})--(\ref{4.3}), with the
potentials of Eqs.~(\ref{4.10}), (\ref{4.11}), and (\ref{4.15}), is
not yet ready to be matched to the black-hole metric of
Eqs.~(\ref{3.1})--(\ref{3.3}). While both metrics are defined in the 
overlap region ${\scr O}$ and describe the same physical situation,
they are expressed in different coordinate systems: The post-Newtonian 
coordinates $(t,x^a)$ are defined everywhere in ${\scr D}$, and they
are attached to the post-Newtonian barycenter; the black-hole
coordinates $(\bar{t},\bar{x}^a)$ are defined only in ${\scr B}$, and
they are attached to the moving black hole. Each coordinate system is
harmonic, and since they overlap in ${\scr O}$ there exists a
coordinate transformation between them. This transformation, from
harmonic coordinates to harmonic coordinates, is worked out in 
Sec.~V A, following closely the treatment of Racine and Flanagan
\cite{racine-flanagan:05}. Their work was built on the post-Newtonian
theory of reference frames developed by Kopeikin, \cite{kopeikin:88},
Brumberg and Kopeikin \cite{brumberg-kopeikin:89}, Damour, Soffel, and
Xu \cite{damour-soffel-xu:91}, and Kopeikin and Vlasov
\cite{kopeikin-vlasov:04}. In Sec.~V B we calculate how the potentials
of Eqs.~(\ref{4.10}), (\ref{4.11}), and (\ref{4.15}) change under the 
transformation from the barycenter frame to the black-hole frame. And
finally, in Sec.~V C we express the transformed potentials in terms of
irreducible quantities that will facilitate the matching with the
black-hole metric, to be carried out in Sec.~VI. 

\subsection{Coordinate transformation} 

The most general coordinate transformation that preserves the
post-Newtonian expansion of the metric is given by
\cite{racine-flanagan:05} 
\begin{eqnarray} 
t &=& \bar{t} + \frac{1}{c^2} \alpha(\bar{t},\bar{x}^a)
+ \frac{1}{c^4} \beta(\bar{t},\bar{x}^a) + O(c^{-6}), 
\label{5.1} \\ 
x^a &=& \bar{x}^a + z^a(\bar{t}) 
+ \frac{1}{c^2} h^a(\bar{t},\bar{x}^a) + O(c^{-4}), 
\label{5.2}
\end{eqnarray} 
where 
\begin{eqnarray} 
\alpha(\bar{t},\bar{x}^a) &=& A(\bar{t}) 
+ v_a \bar{x}^a, 
\label{5.3} \\ 
h^a(\bar{t},\bar{x}^a) &=& H^a(\bar{t}) 
+ H^a_{\ b}(\bar{t}) \bar{x}^b 
+ \frac{1}{2} H^a_{\ bc}(\bar{t}) \bar{x}^b \bar{x}^c, \quad
\label{5.4}
\end{eqnarray}
with 
\begin{eqnarray} 
H_{ab}(\bar{t}) &=& \epsilon_{abc} R^c(\bar{t}) 
+ \frac{1}{2} v_a v_b - \delta_{ab} 
\Bigl( \dot{A} - \frac{1}{2} v^2 \Bigr), \quad
\label{5.5} \\
H_{abc}(\bar{t}) &=& -\delta_{ab} a_c - \delta_{ac} a_b 
+ \delta_{bc} a_a. 
\label{5.6}
\end{eqnarray} 
The functions $A$, $z^a$, $H^a$, and $R^a$ are freely specifiable
functions of time $\bar{t}$, while $\beta$ is a free function of all
the coordinates; these functions characterize the coordinate
transformation. An overdot indicates differentiation with respect to
$\bar{t}$, and we have introduced  
\begin{equation} 
v^a := \dot{z}^a, \qquad a^a := \dot{v}^a = \ddot{z}^a.  
\label{5.7}
\end{equation}
As before indices are raised and lowered with the Euclidean metric
$\delta_{ab}$, and we let $v^2 = \delta_{ab} v^a v^b$. 

The transformation of Eqs.~(\ref{5.1}) and (\ref{5.2}) preserves the
post-Newtonian expansion of the metric, but it does not necessarily
keep the coordinates harmonic. To preserve this also we must set  
\begin{eqnarray} 
\beta(\bar{t},\bar{x}^a) &=& 
\frac{1}{6} \ddot{A} \delta_{ab} \bar{x}^a \bar{x}^b 
\nonumber \\ & & \mbox{} 
+ \frac{1}{30} \bigl( \delta_{ab} \dot{a}_c + \delta_{ac} \dot{a}_b  
+ \delta_{bc} \dot{a}_a \bigr) \bar{x}^a \bar{x}^b \bar{x}^c 
\nonumber \\ & & \mbox{} 
+ \gamma(\bar{t},\bar{x}^a), 
\label{5.8} 
\end{eqnarray} 
and $\gamma$ is required to satisfy Laplace's equation:
$\bar{\nabla}^2 \gamma = 0$, with $\bar{\nabla}^2$ denoting the
Laplacian operator in the coordinates $\bar{x}^a$.  

Under the coordinate transformation the potentials become 
\begin{eqnarray} 
\bar{U}(\bar{t},\bar{x}^a) &=& \hat{U} - \dot{A} 
+ \frac{1}{2} v^2 - a_a \bar{x}^a, 
\label{5.9} \\
\bar{U}^a(\bar{t},\bar{x}^a) &=& \hat{U}^a - \hat{U} v^a 
+ \frac{1}{4} \biggl( V^a + V^a_{\ \ b} \bar{x}^b 
+ \frac{1}{2} V^a_{\ \ bc} \bar{x}^b \bar{x}^c 
\nonumber \\ & & \mbox{}
+ \partial_{\bar{a}} \gamma \biggr), 
\label{5.10} \\ 
\bar{\Psi}(\bar{t},\bar{x}^a) &=& \hat{\Psi} 
- 4 \hat{U}^a v_a + 2 v^2 \hat{U} 
+ \bigl(A + v_b \bar{x}^b\bigr) \partial_{\bar{t}} \hat{U}  
\nonumber \\ & & \mbox{}
+ \Bigl( F^a + F^a_{\ b} \bar{x}^b 
+ \frac{1}{2} F^a_{\ bc} \bar{x}^b \bar{x}^c \Bigr) 
  \partial_{\bar{a}} \hat{U} 
\nonumber \\ & & \mbox{}
+ G + G_a \bar{x}^a + \frac{1}{2} G_{ab} \bar{x}^a \bar{x}^b 
+ \frac{1}{6} G_{abc} \bar{x}^a \bar{x}^b \bar{x}^c 
\nonumber \\ & & \mbox{}
- \partial_{\bar{t}} \gamma,
\label{5.11}
\end{eqnarray} 
where 
\begin{eqnarray} 
V^a &=& (2 \dot{A} - v^2) v^a - \dot{H}^a 
+ \epsilon^a_{\ bc} v^b R^c, 
\label{5.12} \\
V^a_{\ \ b} &=& \frac{3}{2} v^a a_b + \frac{1}{2} a^a v_b 
+ \delta^a_{\ b} \Bigl( \frac{4}{3} \ddot{A} - 2 v^c a_c \Bigr) 
\nonumber \\ & & \mbox{}
- \epsilon^a_{\ bc} \dot{R}^c, 
\label{5.13} \\ 
V^a_{\ \ bc} &=& \frac{6}{5} \bigl( \delta^a_{\ b} \dot{a}_c
+ \delta^a_{\ c} \dot{a}_b \bigr) 
- \frac{4}{5} \delta_{bc} \dot{a}^a, 
\label{5.14} \\ 
F^a &=& H^a - A v^a, 
\label{5.15} \\ 
F^a_{\ b} &=& 
- \delta^a_{\ b} \Bigl( \dot{A} - \frac{1}{2} v^2 \Bigr) 
- \frac{1}{2} v^a v_b + \epsilon^a_{\ bc} R^c, 
\label{5.16} \\ 
F^a_{\ b c} &=& -\bigr( \delta^a_{\ b} a_c + \delta^a_{\ c} a_b \bigr) 
+ a^a \delta_{bc}, 
\label{5.17} \\ 
G &=& \frac{1}{2} \dot{A}^2 - \dot{A} v^2 + \frac{1}{4} v^4 
+ \dot{H}^a v_a, 
\label{5.18} \\
G_a &=& \Bigl(\dot{A} - \frac{1}{2} v^2 \Bigr) a_a
- \Bigl( \ddot{A} - \frac{3}{2} v^c a_c \Bigr) v_a
\nonumber \\ & & \mbox{}
- \epsilon_{abc} v^b \dot{R}^c, 
\label{5.19} \\ 
G_{ab} &=& a_a a_b - v_a \dot{a}_b - \dot{a}_a v_b 
+ \delta_{ab} (v_c \dot{a}^c) 
\nonumber \\ & & \mbox{}
- \frac{1}{3} \delta_{ab} A^{(3)}, 
\label{5.20} \\ 
G_{abc} &=& -\frac{1}{5} \bigl( \delta_{ab} \ddot{a}_c 
+ \delta_{ac} \ddot{a}_b + \delta_{bc} \ddot{a}_a \bigr). 
\label{5.21} 
\end{eqnarray}
Here $A^{(3)}$ stands for $d^3 A/d\bar{t}^3$. 

The ``hatted'' potentials are equal to the original potentials
evaluated at time $t = \bar{t}$ and position 
$x^a = \bar{x}^a + z^a(\bar{t})$. For example, 
\begin{equation} 
\hat{U}(\bar{t},\bar{x}^a) := 
U\bigl(t = \bar{t}, x^a = \bar{x}^a +z^a(\bar{t})\bigr). 
\label{5.22}
\end{equation} 
Because $U$ now possesses, in addition to its original explicit time 
dependence, an implicit time dependence contained in $z^a(\bar{t})$,
some care must be exercised when taking time derivatives of
$\hat{U}$. We have, for example, 
\begin{eqnarray} 
\frac{\partial \hat{U}}{\partial \bar{t}} 
&=& \biggl( \frac{\partial U}{\partial t} 
+ v^a \frac{\partial U}{\partial x^a} 
\biggr)_{t=\bar{t}, x=\bar{x}+z}, 
\nonumber \\ 
& & \label{5.23} \\ 
\frac{\partial \hat{U}}{\partial \bar{x}^a} 
&=& \biggl( \frac{\partial U}{\partial x^a} 
\biggr)_{t=\bar{t}, x=\bar{x}+z}. 
\nonumber 
\end{eqnarray} 
The harmonic coordinate condition reads 
\begin{equation} 
\partial_{\bar{t}} \hat{U} - v^a \partial_{\bar{a}} \hat{U}
+ \partial_{\bar{a}} \hat{U}^a = 0
\label{5.24}
\end{equation} 
when it is expressed in terms of the hatted potentials.

\subsection{Post-Newtonian potentials in the black-hole frame} 

Our task in this subsection is to transform the potentials of
Eqs.~(\ref{4.10}), (\ref{4.11}), and (\ref{4.15}) from the barycenter
coordinates $(t,x^a)$ to coordinates $(\bar{t},\bar{x}^a)$ that are
centered on the black hole. Each coordinate system is harmonic, and
the transformation was described in the preceding subsection. The most
important pieces of the coordinate transformation are the functions
$z^a(\bar{t})$, and for these we choose  
\begin{equation} 
z^a(\bar{t}) = z^a(t = \bar{t}),  
\label{5.25}
\end{equation} 
where $\bm{z}(t)$ is the black hole's position vector in the
barycentric frame. In words, the coordinate displacements
$z^a(\bar{t})$ are given by the black-hole position in the barycentric
system, evaluated at the time $t = \bar{t}$. With this choice it
follows that the quantities $v^a$ and $a^a$ that appear in the
coordinate transformation are the same as those contained in the
post-Newtonian potentials. The other freely specifiable pieces of the
coordinate transformation will be determined in due course.  

The transformed potentials $\bar{U}$, $\bar{U}^a$, and $\bar{\Psi}$
are expressed partly in terms of the hatted potentials. We have, for
example, $\hat{U}(\bar{t},\bar{x}^a) = U(\bar{t},\bar{x}^a +
z^a)$. According to Eqs.~(\ref{4.10}), (\ref{4.11}), and (\ref{4.15}),
the hatted potentials are 
\begin{eqnarray} 
\hat{U} &=& \frac{Gm}{\bar{r}} + \hat{U}_{\rm ext},
\label{5.26} \\ 
\hat{U}^a &=& \frac{Gmv^a}{\bar{r}} + \hat{U}^a_{\rm ext}, 
\label{5.27} \\
\hat{\Psi} &=& -\frac{Gm}{2 \bar{r}^3} v_a v_b \bar{x}^a \bar{x}^b  
+ \frac{Gm}{\bar{r}} \Bigl( \mu + \frac{1}{2} v^2 \Bigr) 
\nonumber \\ & & \mbox{}
- \frac{Gm}{2\bar{r}} a_a \bar{x}^a + \hat{\Psi}_{\rm ext}, 
\label{5.28}
\end{eqnarray} 
in which the vector $\bm{s} := \bm{x} - \bm{z}(t)$ has been identified 
with $\bm{\bar{x}}$; we continue to use the notation $\bar{r} :=
|\bm{\bar{x}}| = (\delta_{ab} \bar{x}^a \bar{x}^b)^{1/2}$. 

The external potentials are given, for example, by   
\begin{equation} 
\hat{U}_{\rm ext}(\bar{t},\bar{x}^a) := 
U_{\rm ext}\bigl(t = \bar{t}, x^a = \bar{x}^a +z^a(\bar{t})\bigr). 
\label{5.29}
\end{equation} 
They satisfy the harmonic coordinate condition 
\begin{equation} 
\partial_{\bar{t}} \hat{U}_{\rm ext} 
- v^a \partial_{\bar{a}} \hat{U}_{\rm ext} 
+ \partial_{\bar{a}} \hat{U}^a_{\rm ext} = 0. 
\label{5.30}
\end{equation} 
Because it is well-behaved near $\bar{x}^a$, each external
potential can be expressed as a Taylor expansion about $\bm{\bar{x}} =  
\bm{0}$. For example,  
\begin{equation} 
\hat{U}_{\rm ext}(\bar{t},\bm{\bar{x}}) = 
\hat{U}_{\rm ext}(\bar{t},\bm{0}) 
+ \partial_{\bar{a}} \hat{U}_{\rm ext} (\bar{t},\bm{0})
+ \frac{1}{2} \partial_{\bar{a}\bar{b}} 
\hat{U}_{\rm ext} (\bar{t},\bm{0}) + \cdots. 
\label{5.31}
\end{equation} 
This defines the strategy behind our calculation of the transformed
potentials: Each quantity that is smooth at $\bm{\bar{x}} = \bm{0}$
will be expressed as a Taylor expansion. The potentials, therefore,
will contain a piece that is singular at $\bm{\bar{x}} = \bm{0}$, and 
a smooth piece that will be expressed as a Taylor series. All Taylor
expansions will be truncated at quadratic order, and they will involve
derivatives of the external potentials evaluated at $\bm{\bar{x}}
= \bm{0}$. 

The harmonic function $\gamma(\bar{t},\bar{x}^a)$, introduced in
Eq.~(\ref{5.8}), is smooth at $\bm{\bar{x}} = \bm{0}$, and it also can 
be expressed as a Taylor expansion. We write  
\begin{eqnarray} 
\gamma(\bar{t},\bar{x}^a) &=& C(\bar{t}) 
+ \gamma_a(\bar{t}) \bar{x}^a 
+ \frac{1}{2} \gamma_{ab}(\bar{t}) \bar{x}^a \bar{x}^b 
\nonumber \\ & & \mbox{}
+ \frac{1}{6} \gamma_{abc}(\bar{t}) \bar{x}^a \bar{x}^b \bar{x}^c 
+ \cdots. 
\label{5.32}
\end{eqnarray} 
To ensure that this is a solution to Laplace's equation, we must
choose the expansion coefficients to be symmetric-tracefree (STF)
tensors. We express this property as $\gamma_{ab} 
= \gamma_{\stf{ab}}$ and $\gamma_{abc} = \gamma_{\stf{abc}}$. The
expansion coefficients are otherwise arbitrary, and they will be
determined in due course.  

After a lengthy computation we obtain 
\begin{eqnarray} 
\bar{U} &=& \frac{Gm}{\bar{r}} + \U{0} + \U{1}_a\, \bar{x}^a  
+ \U{2}_{ab}\, \bar{x}^a \bar{x}^b + \cdots, 
\label{5.33} \\ 
\bar{U}^a &=& \U{0}^a + \U{1}^a_{\ b}\, \bar{x}^b 
+ \U{2}^a_{\ bc}\, \bar{x}^b \bar{x}^c + \cdots, 
\label{5.34} \\ 
\bar{\Psi} &=& 
-\frac{Gm}{\bar{r}^3} \bigl( H_a - A v_a \bigr) \bar{x}^a 
+ \frac{Gm}{\bar{r}} \bigl( \mu + \dot{A} - 2v^2 \bigr) 
\nonumber \\ & & \mbox{}
+ \PPsi{0} + \PPsi{1}_a\, \bar{x}^a 
+ \PPsi{2}_{ab}\, \bar{x}^a \bar{x}^b + \cdots, 
\label{5.35} 
\end{eqnarray} 
with 
\begin{widetext} 
\begin{eqnarray} 
\U{0} &=& \frac{1}{2} v^2 - \dot{A} + \hat{U}_{\rm ext}, 
\label{5.36} \\ 
\U{1}_a &=& -a_a + \partial_{\bar{a}} \hat{U}_{\rm ext}, 
\label{5.37} \\ 
\U{2}_{ab} &=& \frac{1}{2} \partial_{\bar{a}\bar{b}} 
   \hat{U}_{\rm ext}, 
\label{5.38} \\ 
\U{0}^a &=& \hat{U}_{\rm ext} - v^a \hat{U}_{\rm ext} 
+ \frac{1}{4} (2\dot{A} - v^2) - \frac{1}{4} \dot{H}^a 
+ \frac{1}{4} \epsilon^a_{\ bc} v^b R^c + \frac{1}{4} \gamma^a, 
\label{5.39} \\ 
\U{1}^a_{\ b} &=& \partial_{\bar{b}} \hat{U}^a_{\rm ext} 
- v^a \partial_{\bar{b}} \hat{U}_{\rm ext} 
+ \frac{3}{8} v^a a_b + \frac{1}{8} a^a v_b 
+ \frac{1}{4} \delta^a_{\ b} \Bigl( \frac{4}{3} \ddot{A} 
  - 2 v^c a_c \Bigr) - \frac{1}{4} \epsilon^a_{\ bc} \dot{R}^c 
+ \frac{1}{4} \gamma^a_{\ b}, 
\label{5.40} \\ 
\U{2}^a_{\ bc} &=& \frac{1}{2} \partial_{\bar{b}\bar{c}} 
  \hat{U}^a_{\rm ext} - \frac{1}{2} v^a \partial_{\bar{b}\bar{c}} 
  \hat{U}_{\rm ext} 
+ \frac{3}{20} \bigl( \delta^a_{\ b} \dot{a}_c 
  + \delta^a_{\ c} \dot{a}_b \bigr) 
- \frac{1}{10} \dot{a}^a \delta_{bc} + \frac{1}{8} \gamma^a_{\ bc}, 
\label{5.41} \\ 
\PPsi{0} &=& \hat{\Psi}_{\rm ext} - 4 v_a \hat{U}^a_{\rm ext} 
+ 2 v^2 \hat{U}_{\rm ext} + A \partial_{\bar{t}} \hat{U}_{\rm ext} 
+ (H^a - A v^a) \partial_{\bar{a}} \hat{U}_{\rm ext} 
+ \frac{1}{2} \dot{A}^2 - \dot{A} v^2 + \frac{1}{4} v^4 
+ \dot{H}^a v_a - \dot{C}, 
\label{5.42} \\ 
\PPsi{1}_a &=& \partial_{\bar{a}} \hat{\Psi}_{\rm ext} 
- 4 v_b \partial_{\bar{a}} \hat{U}^b_{\rm ext}
+ \Bigl( \frac{5}{2} v^2 - \dot{A} \Bigr) 
   \partial_{\bar{a}} \hat{U}_{\rm ext} 
- \frac{1}{2} v_a v^b \partial_{\bar{b}} \hat{U}_{\rm ext}
+ v_a \partial_{\bar{t}} \hat{U}_{\rm ext} 
+ A \partial_{\bar{t}\bar{a}} \hat{U}_{\rm ext}
\nonumber \\ & & \mbox{}
+ (H^b - A v^b) \partial_{\bar{a}\bar{b}} \hat{U}_{\rm ext}
+ \Bigl( \dot{A} - \frac{1}{2} v^2 \Bigr) a_a 
- \Bigl( \ddot{A} - \frac{3}{2} v^c a_c \Bigr) v_a 
- \epsilon_{abc} \bigl( \partial_{\bar{b}} \hat{U}_{\rm ext} R^c 
  + v^b \dot{R}^c \bigr) - \dot{\gamma}_a, 
\label{5.43} \\
\PPsi{2}_{ab} &=& \frac{1}{2} \partial_{\bar{a}\bar{b}}
   \hat{\Psi}_{\rm ext} 
- 2 v_c \partial_{\bar{a}\bar{b}} \hat{U}^c_{\rm ext} 
+ \Bigl( \frac{3}{2} v^2 - \dot{A} \Bigr) 
  \partial_{\bar{a}\bar{b}} \hat{U}_{\rm ext} 
- \frac{1}{2} v^c v_{(a} \partial_{\bar{b})\bar{c}}
  \hat{U}_{\rm ext} 
+ v_{(a} \partial_{\bar{b})\bar{t}} \hat{U}_{\rm ext} 
+ \frac{1}{2} A \partial_{\bar{t}\bar{a}\bar{b}} \hat{U}_{\rm ext} 
\nonumber \\ & & \mbox{}
- a_{(a} \partial_{\bar{b})} \hat{U}_{\rm ext} 
+ \frac{1}{2} \delta_{ab} a^c \partial_{\bar{c}} \hat{U}_{\rm ext} 
- \epsilon^c_{\ p(a} R^p \partial_{\bar{b})\bar{c}} \hat{U}_{\rm ext} 
+ \frac{1}{2} (H^c - A v^c) \partial_{\bar{c}\bar{a}\bar{b}} 
  \hat{U}_{\rm ext} 
+ \frac{1}{2} a_a a_b - v_{(a} \dot{a}_{b)} 
\nonumber \\ & & \mbox{}
+ \frac{1}{2} \delta_{ab} (v^c \dot{a}_c) 
- \frac{1}{6} \delta_{ab} A^{(3)} 
- \frac{1}{2} \dot{\gamma}_{ab}. 
\label{5.44}
\end{eqnarray} 
It is understood that in these expressions, the external potentials
are evaluated at $\bm{\bar{x}} = \bm{0}$ after differentiation. This
notational convention will be retained below.  
\end{widetext} 

\subsection{Decomposition into irreducible pieces} 

To facilitate the matching procedure it is useful to decompose the
tensors $\U{2}_{ab}$, $\U{1}_{ab}$, $\U{2}_{abc}$, and $\PPsi{2}_{ab}$
into their irreducible components. Equation (\ref{5.38}) reveals that 
$\U{2}_{ab}$ is already a pure STF tensor, because $\bar{\nabla}^2
\hat{U}_{\rm ext} = 0$ everywhere near the black hole. 

We write 
\begin{equation} 
\U{1}_{ab} = \frac{1}{3} \delta_{ab}\, \U{1} + \U{1}_\stf{ab} 
+ \U{1}_{[ab]}, 
\label{5.45}
\end{equation} 
which is a decomposition of $\U{1}_{ab}$ into a trace part, an STF
part, and an antisymmetric part. Equation (\ref{5.40}) implies 
\begin{eqnarray} 
\U{1} &=& \ddot{A} - \partial_{\bar{t}} \hat{U}_{\rm ext} - v^c a_c, 
\label{5.46} \\ 
\U{1}_\stf{ab} &=& 
\partial_{\langle\bar{a}} \hat{U}_{b\rangle}^{\rm ext}   
- v_{\langle a} \partial_{\bar{b} \rangle} \hat{U}_{\rm ext} 
+ \frac{1}{2} v_{\langle a} a_{b \rangle} 
\nonumber \\ & & \mbox{}
+ \frac{1}{4} \gamma_{ab},
\label{5.47} \\ 
\U{1}_{[ab]} &=& -\partial_{[\bar{a}} \hat{U}_{b]}^{\rm ext} 
- v_{[a} \partial_{\bar{b}]} \hat{U}_{\rm ext} 
+ \frac{1}{4} v_{[a} a_{b]} 
\nonumber \\ & & \mbox{}
- \frac{1}{4} \epsilon_{abc} \dot{R}^c. 
\label{5.48}
\end{eqnarray} 

To decompose $\U{2}_{abc}$ we first isolate its completely symmetric
part, and we write $U_{abc} = U_{(abc)} + V_{abc}$, where
$V_{abc} = \frac{2}{3}(U_{[ab]c} + U_{[ac]b})$ is what is left
over of $U_{abc}$ after complete symmetrization. The first term is
decomposed into trace and STF parts. For the second term we note that  
$\frac{2}{3} U_{[ab]c}$ possesses $3 \times 3 = 9$ independent
components, so that it can be expressed as 
$\epsilon_{abp} X^p_{\ c}$, in terms of a general $3 \times 3$ matrix
$X_{ab}$. This, in turn, can be decomposed as $X_{ab} = \frac{1}{3}
\delta_{ab} X + V_{ab} + X_{[ab]}$, in terms of trace, STF, and
antisymmetric components. Finally, we write $X_{[ab]} = \epsilon_{abp}
V^p$, which relates the 3 independent components of $X_{[ab]}$ to
those of a vector $V^a$. Altogether, we have 
$\frac{2}{3} U_{[ab]c} = \frac{1}{3} \epsilon_{abc} X 
+ \delta_{ac} V_b - \delta_{bc} V_a + \epsilon_{abp} V^p_{\ c}$. This
produces $V_{abc} = \delta_{ab} V_{c} + \delta_{ac} V_b 
- 2 \delta_{bc} V_a + \epsilon_{abp} V^p_{\ c} 
+ \epsilon_{acp} V^p_{\ b}$, and we obtain the decomposition 
\begin{eqnarray*} 
\U{2}_{abc} &=& \U{2}_\stf{abc} 
+ \frac{1}{5} \bigl( \delta_{ab}\, \U{2}_c  
+ \delta_{ac}\, \U{2}_b + \delta_{bc}\, \U{2}_a \bigr) 
\\ & & \mbox{}
+ \delta_{ab}\, \V{2}_{c} + \delta_{ac}\, \V{2}_b 
- 2 \delta_{bc}\, \V{2}_a 
\\ & & \mbox{}
+ \epsilon_{abp}\, \V{2}^p_{\ c} 
+ \epsilon_{acp}\, \V{2}^p_{\ b}, 
\end{eqnarray*}
where $\U{2}_a := \delta^{bc} \U{2}_{(abc)}$. The 18 independent
components of $\U{2}_{abc}$ have been packaged into the 7 components
of $\U{2}_\stf{abc}$, the 3 components of $\U{2}_a$, the 3 components
of $\V{2}_a$, and the 5 components of $\V{2}_{ab}$. Calculation shows
that $\V{2}_a = \frac{1}{4} \U{2}_a$ and we obtain, finally, 
\begin{widetext} 
\begin{equation} 
\U{2}_{abc} = \U{2}_\stf{abc} 
+ \frac{9}{20} \bigl( \delta_{ab}\, \U{2}_c  
+ \delta_{ac}\, \U{2}_b \bigr) - \frac{3}{10} \delta_{bc}\, \U{2}_a  
+ \epsilon_{abp}\, \V{2}^p_{\ c} + \epsilon_{acp}\, \V{2}^p_{\ b}, 
\label{5.49}
\end{equation} 
with 
\begin{eqnarray} 
\U{2}_a &=& \frac{1}{3} \bigl( \dot{a}_a 
- \partial_{\bar{t}\bar{a}} \hat{U}_{\rm ext} \bigr), 
\label{5.50} \\ 
\V{2}_{ab} &=& -\frac{1}{6} \epsilon_{(a}^{\ \ pq}
\partial_{\bar{b})\bar{p}} \bigl( \hat{U}^{\rm ext}_q 
- v_q \hat{U}_{\rm ext} \bigr), 
\label{5.51} \\
\U{2}_\stf{abc} &=& \frac{1}{6} \bigl( 
\partial_{\bar{a}\bar{b}} \hat{U}^{\rm ext}_c 
+ \partial_{\bar{a}\bar{c}} \hat{U}^{\rm ext}_b
+ \partial_{\bar{b}\bar{c}} \hat{U}^{\rm ext}_a \bigr) 
- \frac{1}{6} \bigl( v_c \partial_{\bar{a}\bar{b}} \hat{U}_{\rm ext} 
+ v_b \partial_{\bar{a}\bar{c}} \hat{U}_{\rm ext} 
+ v_a \partial_{\bar{b}\bar{c}} \hat{U}_{\rm ext} \bigr) 
\nonumber \\ & & \mbox{}
+ \frac{1}{15} \partial_{\bar{t}} \bigl(
  \delta_{ab} \partial_{\bar{c}} \hat{U}_{\rm ext} 
+ \delta_{ac} \partial_{\bar{b}} \hat{U}_{\rm ext}  
+ \delta_{bc} \partial_{\bar{a}} \hat{U}_{\rm ext} \bigr) 
+ \frac{1}{8} \gamma_{abc}. 
\label{5.52}
\end{eqnarray} 
\end{widetext} 
That the right-hand side of Eq.~(\ref{5.52}) is STF follows from the 
facts that: (i) the potentials $\hat{U}_{\rm ext}$ and 
$\hat{U}^a_{\rm ext}$ satisfy Laplace's equation; (ii) they obey the
harmonic condition of Eq.~(\ref{5.30}); and (iii) $\gamma_{abc}$ is
itself an STF tensor. 

The decomposition of $\PPsi{2}_{ab}$ is 
\begin{widetext} 
\begin{equation} 
\PPsi{2}_{ab} = \PPsi{2}_\stf{ab} 
+ \frac{1}{3} \delta_{ab}\, \PPsi{2},  
\label{5.53}
\end{equation} 
with 
\begin{eqnarray} 
\PPsi{2} &=& \frac{1}{2} \bigl( 
\partial_{\bar{t}\bar{t}} \hat{U}_{\rm ext}   
+ a^2 + v^a \dot{a}_a - A^{(3)} \bigr), 
\label{5.54} \\ 
\PPsi{2}_\stf{ab} &=&   
\frac{1}{2} \partial_{\bar{a}\bar{b}} \hat{\Psi}_{\rm ext} 
- 2 v_c \partial_{\bar{a}\bar{b}} \hat{U}^c_{\rm ext} 
+ \Bigl( \frac{3}{2} v^2 - \dot{A} \Bigr) 
  \partial_{\bar{a}\bar{b}} \hat{U}_{\rm ext} 
- \frac{1}{2} v^c v_{(a} \partial_{\bar{b})\bar{c}}
  \hat{U}_{\rm ext} 
+ v_{(a} \partial_{\bar{b})\bar{t}} \hat{U}_{\rm ext} 
+ \frac{1}{2} A \partial_{\bar{t}\bar{a}\bar{b}} \hat{U}_{\rm ext} 
\nonumber \\ & & \mbox{}
- a_{(a} \partial_{\bar{b})} \hat{U}_{\rm ext} 
+ \frac{1}{2} \delta_{ab} a^c \partial_{\bar{c}} \hat{U}_{\rm ext} 
- \epsilon^c_{\ p(a} R^p \partial_{\bar{b})\bar{c}} \hat{U}_{\rm ext} 
+ \frac{1}{2} (H^c - A v^c) \partial_{\bar{c}\bar{a}\bar{b}} 
  \hat{U}_{\rm ext} 
- \frac{1}{6} \delta_{ab} \partial_{\bar{t}\bar{t}} \hat{U}_{\rm ext} 
\nonumber \\ & & \mbox{}
+ \frac{1}{2} a_{\langle a} a_{b\rangle} - v_{\langle a} \dot{a}_{b\rangle} 
- \frac{1}{2} \dot{\gamma}_{ab}.
\label{5.55}
\end{eqnarray} 
\end{widetext}
The right-hand side of Eq.~(\ref{5.55}) is STF by virtue of the field 
equation satisfied by $\hat{\Psi}_{\rm ext}$. In the barycentric frame
we have $\nabla^2 \Psi_{\rm ext} = 2 \partial_{tt} U_{\rm ext}$;
transforming to the black hole's moving frame gives instead 
\begin{eqnarray} 
\bar{\nabla}^2 \hat{\Psi}_{\rm ext} &=& \partial_{\bar{t}\bar{t}}
\hat{U}_{\rm ext} - a^a \partial_{\bar{a}} \hat{U}_{\rm ext} 
- 2v^a \partial_{\bar{t}\bar{a}} \hat{U}_{\rm ext} 
\nonumber \\ & & \mbox{}
+ v^a v^b \partial_{\bar{a}\bar{b}} \hat{U}_{\rm ext}. 
\label{5.56} 
\end{eqnarray} 

\section{Matching the black-hole and post-Newtonian metrics} 

Because the black-hole and post-Newtonian metrics are now expressed in
the same coordinate system $(\bar{t},\bar{x}^a)$ in the overlap region 
${\scr O}$, we are finally ready to compare their expressions. We
recall that the black-hole metric is valid when $\bar{r} 
\ll {\cal L}$, while the global post-Newtonian metric is valid when
$\bar{r} \gg M$; the expansions provided in
Eqs.~(\ref{5.33})--(\ref{5.35}) are also 
restricted to the domain $\bar{r} \ll {\cal L}$. The metrics can be
compared directly when $\bar{r}$ is such that $M \ll \bar{r} 
\ll {\cal L}$. A typical value of $\bar{r}$ in ${\scr O}$ might be
$\bar{r}_c \sim \sqrt{M {\cal L}}$, and we have that 
$\bar{r}_c/{\cal L} \sim M/\bar{r}_c \sim v/c$; each quantity is
indeed small. We write the metrics in Sec.~VI A in a form that is
ready for matching, and the matching conditions are extracted in
Sec.~VI B. In Sec.~VI C we use them to determine the free
functions associated with the coordinate transformation; this 
procedure reveals, in particular, the equations of motion for the
black hole. In Sec.~VI D we determine the free functions that appear
in the black-hole and post-Newtonian metrics; it is here that the
tidal moments $\bar{\E}_{ab}$ and $\bar{\B}_{ab}$ are finally
obtained. In our last subsection, Sec.~VI E, we transform the
equations of motion and tidal moments from the black-hole frame back
to the barycenter frame, in which they are most easily interpreted.    

\subsection{Metrics} 

In the overlap region ${\scr O}$ the black-hole metric can be
expressed as a post-Newtonian expansion. Writing 
\begin{equation}
\bar{\cal E}_{ab} = \bar{\cal E}^{\mbox{\sc n}}_{ab} 
+ \frac{1}{c^2} \bar{\cal E}^{\mbox{\sc pn}}_{ab} + O(c^{-4}), 
\label{6.1}
\end{equation} 
we get 
\begin{eqnarray}
g_{\bar{0}\bar{0}} &=& 
-1 + \frac{2Gm}{c^2 \bar{r}} - \frac{2G^2m^2}{c^4 \bar{r}^2} 
- \frac{1}{c^2} \bar{\cal E}^{\mbox{\sc n}}_{ab} \bar{x}^a \bar{x}^b 
- \frac{1}{c^4} \bar{\cal E}^{\mbox{\sc pn}}_{ab} \bar{x}^a \bar{x}^b  
\nonumber \\ & & \mbox{}
+ \frac{2Gm}{c^4 \bar{r}} \bar{\cal E}^{\mbox{\sc n}}_{ab} 
  \bar{x}^a \bar{x}^b  
+ O(c^{-6}) + O\biggl(\frac{\bar{r}^3}{{\cal R}^2{\cal L}}\biggr)
\label{6.2}
\end{eqnarray}
and 
\begin{equation} 
g_{\bar{0}\bar{a}} = \frac{2}{3c^3} 
  \epsilon_{abp} \bar{\cal B}^p_{\ c}(\bar{t}) 
  \bar{x}^b \bar{x}^c + O(c^{-5}) 
+ O\biggl(\frac{v}{c}\frac{\bar{r}^3}{{\cal R}^2{\cal L}}\biggr). 
\label{6.3}
\end{equation} 
The post-Newtonian metric is obtained by inserting
Eqs.~(\ref{5.33})--(\ref{5.35}) within
Eqs.~(\ref{4.1})--(\ref{4.3}). The result is 
\begin{widetext} 
\begin{eqnarray} 
g_{\bar{0}\bar{0}} &=& -1 + \frac{2Gm}{c^2\bar{r}} 
- \frac{2Gm}{c^4\bar{r}^3} \bigl( H_a - A v_a \bigr) \bar{x}^a 
+ \frac{2Gm}{c^4\bar{r}} (\mu + \dot{A} - 2v^2) 
- \frac{2G^2m^2}{c^4 \bar{r}^2} 
\nonumber \\ & & \mbox{}
- \frac{4Gm}{c^4 \bar{r}} \bigl( \U{0} + \U{1}_a \bar{x}^a 
  + \U{2}_{ab} \bar{x}^a \bar{x}^b \bigr) 
+ \frac{2}{c^2} \biggl( \U{0} + \frac{1}{c^2} \PPsi{0} 
  - \frac{1}{c^2} (\U{0})^2 \biggr)
\nonumber \\ & & \mbox{}
+ \frac{2}{c^2} \biggl( \U{1}_a + \frac{1}{c^2} \PPsi{1}_a 
  - \frac{2}{c^2} \U{0} \U{1}_a \biggr) \bar{x}^a 
+ \frac{2}{c^2} \biggl( \U{2}_{ab} + \frac{1}{c^2} \PPsi{2}_{ab}  
  - \frac{2}{c^2} \U{0} \U{2}_{ab} 
  - \frac{1}{c^2} \U{1}_a \U{1}_b \biggr) \bar{x}^a \bar{x}^b  
\nonumber \\ & & \mbox{}
+ O(c^{-6}) + O\biggl(\frac{\bar{r}^3}{{\cal R}^2{\cal L}}\biggr)
\label{6.4}
\end{eqnarray}
and 
\begin{equation} 
g_{\bar{0}\bar{a}} = 
-\frac{4}{c^3} \bigl( \U{0}_a + \U{1}_{ab} \bar{x}^b 
+ \U{2}_{abc} \bar{x}^b \bar{x}^c \bigr) + O(c^{-5}) 
+ O\biggl(\frac{v}{c}\frac{\bar{r}^3}{{\cal R}^2{\cal L}}\biggr). 
\label{6.5}
\end{equation} 
\end{widetext}
Comparing Eq.~(\ref{6.4}) to Eq.~(\ref{6.2}), and Eq.~(\ref{6.5}) to 
Eq.~(\ref{6.3}) produces a complete set of matching conditions.  

\subsection{Matching conditions} 

From the absence of $\bar{x}^a$-independent terms in Eq.~(\ref{6.2})
we get $\U{0} + c^{-2}[ \PPsi{0} - (\U{0})^2] = O(c^{-4})$, which
implies that $\U{0} = O(c^{-2})$. The equation simplifies to 
\begin{equation} 
\U{0} + \frac{1}{c^2} \PPsi{0} = O(c^{-4}). 
\label{6.6}
\end{equation}
From the absence of terms linear in $\bar{x}^a$ we get $\U{1}_a +
c^{-2}[ \PPsi{1}_a - 2\U{0} \U{1}_a] = O(c^{-4})$, which implies  
\begin{equation} 
\U{1}_a + \frac{1}{c^2} \PPsi{1}_a = O(c^{-4}). 
\label{6.7}
\end{equation} 
Comparing the singular terms in $g_{\bar{0}\bar{0}}$ and taking into
account the facts that $\U{0} = O(c^{-2})$ and $\U{1} = O(c^{-2})$, we
obtain  
\begin{equation} 
H^a - A v^a = O(c^{-2}), 
\label{6.8}
\end{equation} 
\begin{equation}
\mu + \dot{A} - 2 v^2 = O(c^{-2}), 
\label{6.9}
\end{equation}
and 
\begin{equation} 
\bar{\cal E}^{\mbox{\sc n}}_{ab} = -2\U{2}_{ab}; 
\label{6.10}
\end{equation} 
the last equation can be valid if and only if $\U{2}_{ab}$ is an STF 
tensor, a property that was already established in Sec.~V C. Equation  
(\ref{6.10}) is recovered by matching the terms that are quadratic in 
$\bar{x}^a$ in $g_{\bar{0}\bar{0}}$, and this also reveals that 
\begin{equation} 
\bar{\cal E}^{\mbox{\sc pn}}_{ab} = -2\PPsi{2}_{ab}. 
\label{6.11}
\end{equation} 
We observe that unlike all preceding equations, Eqs.~(\ref{6.10}) and
(\ref{6.11}) do not include an error term $O(c^{-2})$; this is because
each member of these equations is defined so as to possess a specific
post-Newtonian order. 

From the absence of $\bar{x}^a$-independent terms in Eq.~(\ref{6.3})
we get  
\begin{equation} 
\U{0}_a = O(c^{-2}), 
\label{6.12}
\end{equation}
and from the absence of linear terms we obtain
\begin{equation} 
\U{1}_{ab} = O(c^{-2}). 
\label{6.13}
\end{equation}
Finally, matching the quadratic terms in $g_{\bar{0}\bar{a}}$ produces  
\begin{equation} 
\epsilon_{pa(b} \bar{\cal B}^p_{\ c)} = -6 \U{2}_{abc} 
+ O(c^{-2}). 
\label{6.14}
\end{equation} 

The matching conditions of Eqs.~(\ref{6.6})--(\ref{6.14}) allow the 
determination of the quantities 
\[
A, \quad z^a, \quad H^a, \quad R^a, \quad C, \quad 
\gamma_a, \quad \gamma_{ab}, \quad \gamma_{abc}
\]
that appear in the transformation from the barycentric frame $(t,x^a)$
to the black hole's moving frame $(\bar{t},\bar{x}^a)$. They allow
also the determination of the functions 
\[
\mu, \qquad \bar{\cal E}^{\mbox{\sc n}}_{ab}, \qquad  
\bar{\cal E}^{\mbox{\sc pn}}_{ab}, \qquad 
\bar{\cal B}_{ab} 
\]
that appear in the post-Newtonian and black-hole metrics.  

\subsection{Determination of the coordinate transformation} 

From Eq.~(\ref{6.6}) we learn that $\U{0} = O(c^{-2})$, and with the 
expression for $\U{0}$ given in Eq.~(\ref{5.36}), we find that
$A(\bar{t})$ is determined by the differential equation 
\begin{equation} 
\dot{A} = \frac{1}{2} v^2 + \hat{U}_{\rm ext}, 
\label{6.15}
\end{equation} 
where $\hat{U}_{\rm ext} \equiv \hat{U}_{\rm ext}(\bar{t},\bm{0})$ is
the external Newtonian potential evaluated at $\bar{x}^a = 0$. With
$A$ determined, Eq.~(\ref{6.8}) implies 
\begin{equation} 
H^a = A v^a + O(c^{-2}), 
\label{6.16}
\end{equation}
and this determines $H^a(\bar{t})$. 

From Eq.~(\ref{6.7}) we learn that $\U{1}_a = O(c^{-2})$, and with the  
expression for $\U{1}_a$ given in Eq.~(\ref{5.37}), we obtain an
expression for the black hole's acceleration vector: 
\begin{equation} 
a_a = \partial_{\bar{a}} \hat{U}_{\rm ext} + O(c^{-2}), 
\label{6.17}
\end{equation} 
where the external potential is evaluated at $\bar{x}^a = 0$ after
differentiation. This is a Newtonian approximation to the acceleration
vector, and the post-Newtonian corrections will be determined below.  

With $\U{0} = O(c^{-2})$ Eq.~(\ref{6.6}) implies $\PPsi{0} =
O(c^{-2})$, and taking into account Eqs.~(\ref{6.15})--(\ref{6.17}),
Eq.~(\ref{5.42}) reveals that  
\begin{eqnarray} 
\dot{C} &=& \hat{\Psi}_{\rm ext} - 4 v_a \hat{U}^a_{\rm ext} 
+ \frac{5}{2} v^2 \hat{U}_{\rm ext} + \frac{1}{2} \hat{U}^2_{\rm ext} 
\nonumber \\ & & \mbox{}
+ A \bigl( \partial_{\bar{t}} \hat{U}_{\rm ext} 
+ v^a \partial_{\bar{a}} \hat{U}_{\rm ext} \bigr)  
+ \frac{3}{8} v^4 + O(c^{-2}), \quad \qquad
\label{6.18}
\end{eqnarray}  
in which each external potential is evaluated at $\bar{x}^a = 0$ after 
differentiation. This equation determines $C(\bar{t})$. 

Equation (\ref{6.13}) implies that each irreducible piece of
$\U{1}_{ab}$ must vanish to order $c^0$. According to Eq.~(\ref{5.45})  
we must have $\U{1} = 0 = \U{1}_\stf{ab} = \U{1}_{[ab]}$. With
Eq.~(\ref{5.46}) we reproduce Eq.~(\ref{6.15}). From Eqs.~(\ref{5.47})
and (\ref{6.17}) we get 
\begin{equation} 
\gamma_{ab} = -4 \partial_{\langle \bar{a}} 
  \hat{U}^{\rm ext}_{b\rangle} + 2 v_{\langle a} 
  \partial_{\bar{b}\rangle} \hat{U}_{\rm ext}. 
\label{6.19}
\end{equation}  
And from Eq.~(\ref{5.48}) we find 
\begin{equation} 
\epsilon_{abc} \dot{R}^c = -4 \partial_{[\bar{a}} 
  \hat{U}^{\rm ext}_{b]} -3 v_{[a} 
  \partial_{\bar{b}]} \hat{U}_{\rm ext}, 
\label{6.20}
\end{equation} 
an equation that determines $R^a(\bar{t})$. 

Taking into account Eqs.~(\ref{6.15})--(\ref{6.17}), Eqs.~(\ref{6.12})
and (\ref{5.39}) imply
\begin{eqnarray} 
\gamma_a &=& -4 \hat{U}_a^{\rm ext} + \Bigl( \frac{1}{2} v^2 
+ 3 \hat{U}_{\rm ext} \Bigr) v_a 
+ A \partial_{\bar{a}} \hat{U}_{\rm ext} 
\nonumber \\ & & \mbox{}
- \epsilon_{abc} v^b R^c + O(c^{-2}). 
\label{6.21}
\end{eqnarray} 

We may now determine the post-Newtonian corrections to the
acceleration vector. We return to Eq.~(\ref{6.7}), in which we insert 
Eqs.~(\ref{5.37}) and (\ref{5.43}). We next incorporate
Eqs.~(\ref{6.15})--(\ref{6.17}), as well as Eqs.~(\ref{6.20}) and 
(\ref{6.21}). After simplification we obtain 
\begin{eqnarray} 
a^a &=& \partial^{\bar{a}} \hat{U}^{\rm ext} 
+ \frac{1}{c^2} \biggl[ \partial^{\bar{a}} \hat{\Psi}_{\rm ext} 
- 4 v_b \partial^{\bar{a}} \hat{U}^b_{\rm ext} 
+ 4 \partial_{\bar{t}} \hat{U}^a_{\rm ext} 
\nonumber \\ & & \mbox{}
+ \bigl( v^2 - 4 \hat{U}_{\rm ext} \bigr) 
  \partial^{\bar{a}} \hat{U}_{\rm ext} 
- \bigl( v^c \partial_{\bar{c}} \hat{U}_{\rm ext} 
+ 3 \partial_{\bar{t}} \hat{U}_{\rm ext} \bigr) v^a \biggl]
\nonumber \\ & & \mbox{}
+ O(c^{-4}), 
\label{6.22}
\end{eqnarray}
where (as always) the external potentials are evaluated at
$\bar{x}^a=0$ after differentiation. Equation (\ref{6.22}) is a system
of second-order differential equations for the functions
$z^a(\bar{t})$; they represent {\it equations of motion} for the black 
hole.   

The last piece of the coordinate transformation that must be
determined is $\gamma_{abc}$. The information comes from
Eq.~(\ref{6.14}) and the decomposition of Eq.~(\ref{5.49}). Comparing
the equations reveals that $\U{2}_a$ and $\U{2}_\stf{abc}$ must both
vanish. The first statement reproduces Eq.~(\ref{6.17}), while the
second implies 
\begin{eqnarray*} 
\gamma_{abc} &=& -\frac{4}{3} \bigl( 
\partial_{\bar{a}\bar{b}} \hat{U}^{\rm ext}_c 
+ \partial_{\bar{a}\bar{c}} \hat{U}^{\rm ext}_b
+ \partial_{\bar{b}\bar{c}} \hat{U}^{\rm ext}_a \bigr) 
\nonumber \\ & & \mbox{}
+ \frac{4}{3} \bigl( v_c \partial_{\bar{a}\bar{b}} \hat{U}^{\rm ext}  
+ v_b \partial_{\bar{a}\bar{c}} \hat{U}^{\rm ext} 
+ v_a \partial_{\bar{b}\bar{c}} \hat{U}^{\rm ext} \bigr) 
\nonumber \\ & & \mbox{}
- \frac{8}{15} \partial_{\bar{t}} \bigl(
  \delta_{ab} \partial_{\bar{c}} \hat{U}_{\rm ext} 
+ \delta_{ac} \partial_{\bar{b}} \hat{U}_{\rm ext}  
+ \delta_{bc} \partial_{\bar{a}} \hat{U}_{\rm ext} \bigr). 
\end{eqnarray*} 
This can be expressed more compactly as 
\begin{equation} 
\gamma_{abc} = -4 \bigl( 
\partial_{\langle\bar{a}\bar{b}} \hat{U}^{\rm ext}_{c\rangle} 
- v_{\langle a}\partial_{\bar{b}\bar{c}\rangle} \hat{U}^{\rm ext}   
\bigr). 
\label{6.23} 
\end{equation} 
The coordinate transformation is now completely determined by the
matching conditions.  

\subsection{Determination of the metric functions}

Equations (\ref{6.9}) and (\ref{6.15}) imply
\begin{equation} 
\mu = \frac{3}{2} v^2 - \hat{U}_{\rm ext}, 
\label{6.24}
\end{equation}
in which the external potential is evaluated at $\bar{x}^a=0$. As was
first pointed out in the footnote that follows Eq.~(\ref{4.13}), this  
assignment can also be obtained by calculating the post-Newtonian
potential $\psi$ --- see Eq.~(\ref{4.4}) --- for a point particle of
the same mass as the black hole. Our matching procedure shows that
$\mu$ keeps the same value when the particle is replaced by the black
hole.  

The electric components of the tidal fields are determined by
Eqs.~(\ref{6.10}) and (\ref{6.11}). After inserting Eqs.~(\ref{5.38})
and (\ref{5.53}) and invoking Eq.~(\ref{6.15}) to eliminate the trace
part of $\PPsi{2}_{ab}$, we obtain 
\begin{equation} 
\bar{\cal E}^{\mbox{\sc n}}_{ab} = 
-\partial_{\bar{a}\bar{b}} \hat{U}_{\rm ext} 
\label{6.25}
\end{equation} 
and $\bar{\cal E}^{\mbox{\sc pn}}_{ab} = -2\PPsi{2}_\stf{ab}$. An
explicit evaluation of this yields 
\begin{widetext} 
\begin{eqnarray} 
\bar{\cal E}^{\mbox{\sc pn}}_{ab} &=& 
-\partial_\stf{\bar{a}\bar{b}} \hat{\Psi}_{\rm ext} 
+ 4 v_c \partial_{\bar{a}\bar{b}} \hat{U}^c_{\rm ext} 
- 4 \partial_{\bar{t} \langle \bar{a}} \hat{U}^{\rm ext}_{b\rangle} 
- 2\bigl( v^2 - \hat{U}_{\rm ext} \bigr) 
  \partial_{\bar{a}\bar{b}} \hat{U}_{\rm ext}
+ v^c v_{\langle a} \partial_{\bar{b}\rangle \bar{c}} 
  \hat{U}_{\rm ext}  
+ 2 v_{\langle a} \partial_{\bar{b}\rangle \bar{t}} 
  \hat{U}_{\rm ext}
\nonumber \\ & & \mbox{} 
+ 3 \partial_{\langle \bar{a}} \hat{U}_{\rm ext} 
  \partial_{\bar{b} \rangle} \hat{U}_{\rm ext} 
- A \partial_{\bar{t}\bar{a}\bar{b}} \hat{U}_{\rm ext} 
+ 2 \epsilon_{cp(a} R^p \partial^{\bar{c}}_{\ \bar{b})} 
  \hat{U}_{\rm ext}, 
\label{6.26}
\end{eqnarray}
\end{widetext}
where, as usual, the external potentials are evaluated at $\bar{x}^a 
= 0$ after differentiation. The complete tidal potentials are 
$\bar{\cal E}_{ab} = \bar{\cal E}^{\mbox{\sc n}}_{ab} + c^{-2} 
\bar{\cal E}^{\mbox{\sc pn}}_{ab} + O(c^{-4})$, as was expressed in 
Eq.~(\ref{6.1}). 

The magnetic components of the tidal fields are determined by
Eq.~(\ref{6.14}) and the decomposition of Eq.~(\ref{5.49}). Taking
into account the facts that $\U{2}_a$ and $\U{2}_\stf{abc}$ must both 
vanish (as was noted previously), comparing the equations reveals that
$\bar{\cal B}_{ab} = -12 \V{2}_{ab} + O(c^{-2})$. With
Eq.~(\ref{5.51}), this is  
\begin{equation} 
\bar{\cal B}_{ab} = 2 \epsilon_{(a}^{\ \ pq}
\partial_{\bar{b})\bar{p}} \bigl( \hat{U}^{\rm ext}_q 
- v_q \hat{U}_{\rm ext} \bigr) + O(c^{-2}). 
\label{6.27}
\end{equation} 
The metric functions are now fully determined by the matching
conditions. 

\subsection{Transformation to the barycentric frame} 

In the moving frame $(\bar{t},\bar{x}^a)$ the black hole is situated
at $\bar{x}^a = 0$. According to Eqs.~(\ref{5.1})--(\ref{5.4}), the
position of the black hole in the barycentric frame is described by 
the parametric equations 
\begin{eqnarray} 
t_{\rm bh} &=& \bar{t} + \frac{1}{c^2} A(\bar{t}) + O(c^{-4}),
\nonumber \\ 
& & \\  
x^a_{\rm bh} &=& z^a(\bar{t}) + \frac{1}{c^2} H^a(\bar{t}) 
+ O(c^{-4}).
\nonumber  
\label{6.28}
\end{eqnarray} 
The first equation can be approximately inverted as $\bar{t} 
= t_{\rm bh} - c^{-2} A(t_{\rm bh}) + O(c^{-4})$, and substitution
into the second equation yields $x^a_{\rm bh} = z^a(t_{\rm bh}) 
+ c^{-2}[ H^a(t_{\rm bh}) - A(t_{\rm bh}) v^a(t_{\rm bh})] 
+ O(c^{-4})$. With Eq.~(\ref{6.16}) this becomes 
\begin{equation} 
x^a_{\rm bh} = z^a(t_{\rm bh}) + O(c^{-4}),
\label{6.29}
\end{equation} 
which is the same statement as Eq.~(\ref{5.25}). The position of the 
black hole in the barycenter frame is therefore obtained simply by
evaluating the functions $z^a(\bar{t})$ at the time
$\bar{t} = t_{\rm bh}$. From this observation it follows that the
black hole's barycentric velocity is $v^a(t_{\rm bh})$, and its
acceleration is $a^a(t_{\rm bh})$. Henceforth we shall omit the label 
``bh'' on the barycentric time coordinate.  

The equations of motion for the black hole, expressed in the
barycentric frame, are obtained from Eq.~(\ref{6.22}) by replacing the 
hatted potentials ($\hat{U}_{\rm ext}$ and so on) with the original
potentials ($U_{\rm ext}$ and so on) using the correspondence of
Eqs.~(\ref{5.22}) and (\ref{5.23}). Noting that Eq.~(\ref{6.22}) is to
be evaluated at $\bar{t} = t_{\rm bh} \equiv t$, we get 
\begin{eqnarray} 
a^a &=& \partial^a U^{\rm ext} 
+ \frac{1}{c^2} \biggl[ \partial^a \Psi_{\rm ext} 
- 4 \bigl( \partial^a U^b_{\rm ext} - \partial^b U^a_{\rm ext} \bigr)
  v_b 
\nonumber \\ & & \mbox{}
+ 4 \partial_t U^a_{\rm ext} 
+ \bigl( v^2 - 4 U_{\rm ext} \bigr) \partial^a U_{\rm ext} 
\nonumber \\ & & \mbox{}
- v^a \bigl( 4 v^b \partial_b U_{\rm ext} 
+ 3 \partial_t U_{\rm ext} \bigr) \biggl] + O(c^{-4}). 
\label{6.30}
\end{eqnarray}
The external potentials were introduced in
Eqs.~(\ref{4.10}), (\ref{4.11}), and (\ref{4.15}), and here they are
evaluated at $\bm{x} = \bm{z}(t)$ after differentiation. Equation
(\ref{6.30}) applies to a black hole moving in any post-Newtonian
environment. When this environment consists of $(N-1)$ external
bodies, so that the black hole is a member of an $N$-body system,
Eq.~(\ref{6.30}) reduces to the standard (Einstein-Infeld-Hoffman)
post-Newtonian equations of motion. Because this connection is well
understood, we shall not provide here a derivation of this well-known
fact; the EIH equations are listed, for example, in Exercise 39.15 of
Misner, Thorne, and Wheeler \cite{MTW:73}.  

Following Racine and Flanagan \cite{racine-flanagan:05} we define
barycentric tidal moments ${\cal E}_{ab}(t)$ and ${\cal B}_{ab}(t)$
that are related to those of the black-hole frame by the
transformation  
\begin{eqnarray} 
{\cal E}_{ab}(t) &:=& 
{\cal M}_a^{\ c}(\bar{t}) {\cal M}_b^{\ d}(\bar{t}) 
  \bar{\cal E}_{cd}(\bar{t}), 
\nonumber \\  
& & \\ 
{\cal B}_{ab}(t) &:=& 
{\cal M}_a^{\ c}(\bar{t}) {\cal M}_b^{\ d}(\bar{t}) 
  \bar{\cal B}_{cd}(\bar{t}),
\nonumber 
\label{6.31}
\end{eqnarray} 
where 
\begin{equation} 
{\cal M}_{ab}(\bar{t}) := \delta_{ab} 
+ \frac{1}{c^2} \epsilon_{abc} R^c(\bar{t}) + O(c^{-4}) 
\label{6.32}
\end{equation}
is a post-Newtonian rotation matrix that accounts for the precession
of the moving frame relative to the barycentric frame. We recall that
the time coordinates are related by $t = \bar{t} + c^{-2} A(\bar{t}) +
O(c^{-4})$. The quantities $A(\bar{t})$ and $R^a(\bar{t})$ that appear
in the transformation are determined by Eqs.~(\ref{6.15}) and
(\ref{6.20}), respectively. The inverse transformation is 
\begin{eqnarray} 
\bar{\cal E}_{ab}(\bar{t}) &=& 
{\cal N}_a^{\ c}(t) {\cal N}_b^{\ d}(t) 
  {\cal E}_{cd}(t), 
\nonumber \\ 
& & \\ 
\bar{\cal B}_{ab}(\bar{t}) &=&
 {\cal N}_a^{\ c}(t) {\cal N}_b^{\ d}(t) 
  {\cal B}_{cd}(t), 
\nonumber 
\label{6.33}
\end{eqnarray} 
where 
\begin{equation} 
{\cal N}_{ab}(t) := \delta_{ab} 
- \frac{1}{c^2} \epsilon_{abc} R^c(t) + O(c^{-4}) 
\label{6.34}
\end{equation}
is the inverse to ${\cal M}_{ab}$. In these equations we have $\bar{t}
= t - c^{-2} A(t) + O(c^{-4})$, and the quantities $A(t)$ and $R^a(t)$
are determined by 
\begin{eqnarray} 
\frac{dA}{dt} &=& \frac{1}{2} v^2 + U_{\rm ext}, 
\nonumber \\  
& & \\
\epsilon_{abc} \frac{dR^c}{dt} &=& -4 \partial_{[a} 
  U^{\rm ext}_{b]} -3 v_{[a} \partial_{b]} U_{\rm ext}. 
\nonumber 
\label{6.35}
\end{eqnarray} 

Expanding Eqs.~(\ref{6.31}) in powers of $c^{-2}$ produces 
${\cal E}_{ab} = \bar{\cal E}_{ab} + c^{-2}[ -A \partial_{\bar{t}}
  \bar{\cal E}_{ab} + 2 \epsilon_{cp(a} R^p \bar{\cal E}^c_{\ b)}] 
+ O(c^{-4})$ and ${\cal B}_{ab} = \bar{\cal B}_{ab} + O(c^{-2})$. In
this we substitute Eqs.~(\ref{6.25})--(\ref{6.27}), and we replace the 
hatted potentials by the original potentials using the correspondence
of Eqs.~(\ref{5.22}) and (\ref{5.23}). After simplification we obtain 
\begin{widetext} 
\begin{eqnarray} 
{\cal E}_{ab} &=& -\partial_{ab} U_{\rm ext} + \frac{1}{c^2} 
\left\lgroup 
-\partial_\stf{ab} \Psi_{\rm ext} 
+ 4v^c \bigl( \partial_{ab} U^{\rm ext}_c 
- \partial_{c\langle a} U^{\rm ext}_{b\rangle} \bigr) 
- 4 \partial_{t\langle a} U^{\rm ext}_{b\rangle} 
- 2(v^2 - U_{\rm ext}) \partial_{ab} U_{\rm ext} 
\right. \nonumber \\ & & \mbox{} \left.
+ 3 v^c v_{\langle a} \partial_{b\rangle c} U_{\rm ext} 
+ 2 v_{\langle a} \partial_{b\rangle t} U_{\rm ext} 
+ 3 \partial_{\langle a} U_{\rm ext} \partial_{b\rangle} U_{\rm ext} 
\right\rgroup + O(c^{-4}) 
\label{6.36}
\end{eqnarray}
\end{widetext} 
and 
\begin{equation} 
{\cal B}_{ab} = 2 \epsilon_{pq(a} \partial^p_{\ b)} \bigl( 
U^q_{\rm ext} - v^q U_{\rm ext} \bigr) + O(c^{-2}). 
\label{6.37}
\end{equation} 
In these equations the external potentials are evaluated at $\bm{x} = 
\bm{z}(t)$ after differentiation. Notice that unlike 
$\bar{\cal E}_{ab}$, the barycentric tidal moment ${\cal E}_{ab}$ does  
not involve the functions $A$ and $R^a$ that must be obtained by
integrating first-order differential equations; this was the reason
for introducing the transformation of Eq.~(\ref{6.31}). Notice also 
that since $\bar{\cal B}_{ab}$ has been worked out to leading-order
only, its transformation to the barycentric frame is trivial. 

\section{Tidal moments for a two-body system} 

The results obtained in the preceding section apply to any
post-Newtonian environment described by the external potentials
$U_{\rm ext}$, $U^a_{\rm ext}$, and $\Psi_{\rm ext}$. Given these
potentials, the motion of the black hole in the barycentric frame
$(t,x^a)$ is determined by Eq.~(\ref{6.30}), and the barycentric tidal
moments $\E_{ab}$ and $\B_{ab}$ are obtained by evaluating
Eqs.~(\ref{6.36}) and (\ref{6.37}), respectively. The tidal moments
perceived by the black hole are $\bar{\E}_{ab}$ and $\bar{\B}_{ab}$,
and these are calculated using the transformation of
Eqs.~(\ref{6.33})--(\ref{6.35}). 

In this section we specialize our discussion to a specific
post-Newtonian environment that consists of a single external body
(perhaps another black hole), so that the black hole is a member of a
post-Newtonian two-body system. Adapting our notation to this specific
situation, we let our original black hole have a mass $m_1$, position
$\bm{z}_1(t)$, velocity $\bm{v}_1(t)$, acceleration $\bm{a}_1(t)$, and
so on. (These quantities were previously denoted $m$, $\bm{z}$,
$\bm{v}$, and $\bm{a}$, respectively. Below we will redefine $m$ to be 
the system's total mass $m_1 + m_2$, and $\bm{v}$ to be the system's 
relative velocity $\bm{v}_1 - \bm{v}_2$.) The second body also is
modeled as a post-Newtonian monopole, and it has a mass $m_2$,
position $\bm{z}_2(t)$, velocity $\bm{v}_2(t)$, acceleration
$\bm{a}_2(t)$, and so on. 

In Sec.~VII A we list the potentials associated with the external
body, evaluate their derivatives, and calculate the barycentric tidal
moments. In Sec.~VII B we simplify our expressions by writing them in
terms of $\bm{r} := \bm{z}_1 - \bm{z}_2$, the relative separation
between the two bodies, from which the individual trajectories can be
recovered. In Sec.~VII C we restrict our attention to a binary system
in circular motion, and in Sec.~VII D we compute the tidal moments as
viewed in the moving frame of the black hole. Finally, in Sec.~VII
E we compare our post-Newtonian answers to those obtained by Poisson
\cite{poisson:04a} in the context of the small-hole approximation (see
Sec.~I A).  

Our calculations in this section rely on well-known results from
post-Newtonian theory. These can be found, for example, in Blanchet's
review article \cite{blanchet:06}. 

\subsection{Two-body potentials and tidal moments} 

Assuming that the external body is a post-Newtonian monopole of mass
$m_2$, the external potentials can be expressed in a form that is
directly analogous to that of the black-hole potentials of
Eqs.~(\ref{4.10})--(\ref{4.13}). We have  
$U_{\rm ext} = Gm_2/s$, $U^a_{\rm ext} = Gm_2 v^a_2/s$, 
$\psi_{\rm ext} = Gm_2\mu_2/s$, $X_{\rm ext} = Gm_2 s$, and 
$\Psi_{\rm ext} = \psi_{\rm ext} + \frac{1}{2} \partial^2_{t}
X_{\rm ext}$, where $s$ now stands for $|\bm{x}-\bm{z}_2(t)|$ and   
$\mu_2 := \frac{3}{2} v_2^2 - Gm_1/|\bm{z}_1-\bm{z}_2|$. 

These potentials are easily differentiated, and after evaluation at
$\bm{x} = \bm{z}_1(t)$ we obtain 
\begin{eqnarray} 
U_{\rm ext} &=& \frac{Gm_2}{r}, 
\label{7.1} \\ 
\partial_a U_{\rm ext} &=& -\frac{Gm_2}{r^2} n_a, 
\label{7.2} \\
\partial_{ab} U_{\rm ext} &=& \frac{Gm_2}{r^3} 
  \bigl( 3n_a n_b - \delta_{ab} \bigr), 
\label{7.3} \\
\partial_{ta} U_{\rm ext} &=& -\frac{Gm_2}{r^3} 
  \bigl( 3n_a n_b - \delta_{ab} \bigr) v_2^b, 
\label{7.4} \\ 
\partial_{b} U^a_{\rm ext} &=& -\frac{Gm_2 v_2^a}{r^2} n_b, 
\label{7.5} \\  
\partial_{bc} U^a_{\rm ext} &=& \frac{Gm_2 v_2^a}{r^3} 
  \bigl( 3n_b n_c - \delta_{bc} \bigr), 
\label{7.6} \\
\partial_{tb} U^a_{\rm ext} &=& -\frac{G^2 m_1 m_2}{r^4} n^a n_b  
\nonumber \\ & & \mbox{}
- \frac{Gm_2 v_2^a}{r^3} \bigl( 3n_b n_c - \delta_{bc} \bigr) v_2^c, 
\label{7.7} \\ 
\partial_{ab} \Psi_{\rm ext} &=& 
\frac{Gm_2}{r^3} \Bigl( 2v_2^2 - \frac{Gm_1}{r} \Bigr) 
    \bigl( 3n_a n_b - \delta_{ab} \bigr)
\nonumber \\ & & \mbox{}
+ \frac{Gm_2}{2r^3} \Bigl[ 3 (\bm{n} \cdot \bm{v}_2)^2 (\delta_{ab} 
  - 5 n_a n_b) 
\nonumber \\ & & \mbox{}
+ 12 (\bm{n} \cdot \bm{v}_2) v_{2(a} n_{b)} 
  - 2v_{2a} v_{2b} \Bigr] 
\nonumber \\ & & \mbox{}
+ \frac{G^2 m_1 m_2}{2 r^4} \bigl( \delta_{ab} - n_a n_b \bigr). 
\label{7.8}
\end{eqnarray} 
We have introduced the new quantities 
\begin{equation} 
\bm{r} := \bm{z}_1 - \bm{z}_2, \qquad 
r := |\bm{z}_1 - \bm{z}_2|, \qquad 
\bm{n} := \bm{r}/r. 
\label{7.9}
\end{equation}
To arrive at Eqs.~(\ref{7.7}) and (\ref{7.8}) we used the equations of   
motion $\bm{a}_2 = G m_1 \bm{n}/r^2 + O(c^{-2})$ to replace the 
acceleration vector of the second body by its Newtonian expression.  

Making the substitutions into Eqs.~(\ref{6.36}) and (\ref{6.37}) gives 
\begin{widetext} 
\begin{eqnarray} 
{\cal E}_{ab} &=& -\frac{3Gm_2}{r^3} n_\stf{ab} 
- \frac{3Gm_2}{c^2 r^3} \left\lgroup \Bigl[ 2v_1^2 
- 4 (\bm{v}_1 \cdot \bm{v}_2) + 2v_2^2 
- \frac{5}{2} (\bm{n} \cdot \bm{v}_2)^2 
- \frac{5}{2} \frac{Gm_1}{r} - 3 \frac{Gm_2}{r} \Bigr] n_\stf{ab} 
\right. \nonumber \\ & & 
- \bigl[ 3 (\bm{n} \cdot \bm{v}_1) - 2 (\bm{n} \cdot \bm{v}_2) \bigr] 
  n_{\langle a} v_{1 b \rangle} 
+ \bigl[ 4 (\bm{n} \cdot \bm{v}_1) - 2 (\bm{n} \cdot \bm{v}_2) \bigr] 
  n_{\langle a} v_{2 b \rangle} 
\nonumber \\ & & \left.
+ v_{1\langle a} v_{1 b\rangle} - 2 v_{1\langle a} v_{2 b\rangle} 
+ v_{2\langle a} v_{2 b\rangle} \right\rgroup 
+ O(c^{-4})
\label{7.10} 
\end{eqnarray}
\end{widetext} 
and 
\begin{equation} 
{\cal B}_{ab} = -\frac{6Gm_2}{r^3} \bigl[ \bm{n} \times (\bm{v}_1 -
  \bm{v}_2) \bigr]_{(a} n_{b)} + O(c^{-2}), 
\label{7.11}
\end{equation}
where $n_\stf{ab} := n_a n_b - \frac{1}{3} \delta_{ab}$. The
quantities $A(t)$ and $R^a(t)$ that appear in the transformation from
the barycentric frame to the black-hole frame are determined by the
equations  
\begin{eqnarray} 
\dot{A} &=& \frac{1}{2} v_1^2 + \frac{Gm_2}{r} + O(c^{-2}),
\nonumber \\ 
& & \label{7.12} \\ 
\dot{\bm{R}} &=& \frac{Gm_2}{2r^2} \bm{n} \times 
  (4 \bm{v}_2 - 3\bm{v}_1) + O(c^{-2}); 
\nonumber 
\end{eqnarray}   
these are obtained from Eqs.~(\ref{6.35}). 

It is straightforward to generalize Eqs.~(\ref{7.10})--(\ref{7.12})
from a two-body system to an $N$-body system by simply writing the
external potentials as a sum of single-body terms. The generalized
expressions can then be compared with the corresponding results of  
Damour, Soffel, and Xu --- see, in particular, Eqs.~(4.29)--(4.31) of
Ref.~\cite{damour-soffel-xu:93}. As was already stated in Sec.~I B, we
find that our expressions agree with theirs. 

\subsection{Generic orbital motion} 

To simplify the foregoing results we incorporate the fact that the
motion of each body in a post-Newtonian two-body system can be related 
to the motion of the {\it relative orbit}, which is described by the
separation vector $\bm{r} = \bm{z}_1 - \bm{z}_2$ and the relative
velocity vector $\bm{v} = \bm{v}_1 - \bm{v}_2$. The post-Newtonian
dynamics implies that if the system's barycenter is placed at the
origin of the coordinate system, then $\bm{z}_1 = (m_2/m) \bm{r} 
+ O(c^{-2})$ and $\bm{z}_2 = -(m_1/m) \bm{r} + O(c^{-2})$, where 
$m := m_1 + m_2$ is the total mass of the two-body system. As a
consequence we also have $\bm{v}_1 = (m_2/m) \bm{v} + O(c^{-2})$ and 
$\bm{v}_2 = -(m_1/m) \bm{v} + O(c^{-2})$, and we make these
substitutions in Eqs.~(\ref{7.10})--(\ref{7.12}).  

In addition we incorporate the fact that the post-Newtonian motion of
a two-body system takes place in a fixed orbital plane. We take this
plane to be the $x$-$y$ plane, and we use polar coordinates $r$ and
$\phi$ to describe the orbital motion. We write $\bm{r} =
(r\cos\phi, r\sin\phi, 0)$, and we resolve all vectors in the basis
$\bm{n} = (\cos\phi,\sin\phi,0)$, $\bm{\phi} =
(-\sin\phi,\cos\phi,0)$, and $\bm{l} = (0,0,1)$ associated with the
polar coordinates; the vector $\bm{l}$ is normal to the plane, and it
is aligned with the system's total angular momentum. We have 
\begin{equation}
\bm{r} = r \bm{n}, \qquad  
\bm{v} = \dot{r} \bm{n} + (r\dot{\phi}) \bm{\phi},
\label{7.13} 
\end{equation} 
and we also make these substitutions in
Eqs.~(\ref{7.10})--(\ref{7.12}). 

After simplification our results become 
\begin{eqnarray} 
{\cal E}_{ab} &=& -\frac{3Gm_2}{r^3} n_\stf{ab} 
- \frac{3Gm_2}{c^2 r^3} \left\lgroup \biggl[ 
-\frac{3m_1^2}{2m^2} \dot{r}^2 
+ 2(r\dot{\phi})^2 
\right. \nonumber \\ & & \mbox{} \left.
- \frac{5Gm_1+6Gm_2}{2r} \biggr] n_\stf{ab} 
\right. \nonumber \\ & & \mbox{} \left.
- \frac{(2m_1+m_2)m_2}{m^2} \dot{r} (r\dot{\phi}) n_{(a} \phi_{b)} 
\right. \nonumber \\ & & \mbox{} \left.
+ (r\dot{\phi})^2 \phi_\stf{ab} \right\rgroup + O(c^{-4}), 
\label{7.14} \\
{\cal B}_{ab} &=& -\frac{6 G m_2}{r^3} (r\dot{\phi})\, l_{(a} n_{b)}  
+ O(c^{-2}), 
\label{7.15} \\
\dot{A} &=& \frac{m_2^2}{2m^2} \bigl[ \dot{r}^2 + (r\dot{\phi})^2 \bigr] 
+ \frac{Gm_2}{r} + O(c^{-2}), 
\label{7.16} \\ 
\dot{R}^a &=& -\frac{G m_2}{2r^2} \frac{4m_1+3m_2}{m}
(r\dot{\phi})\, l^a + O(c^{-2}). 
\label{7.17}
\end{eqnarray} 
We recall that $m = m_1 + m_2$ is the total mass of the system,
$r := |\bm{z}_1 - \bm{z}_2|$ is the inter-body distance, $\phi$
is the angular position of the relative orbit in the orbital plane,
$\bm{n}$ is a unit vector that points from body 2 to body 1,
$\bm{\phi}$ is a unit vector that points in the direction of
increasing $\phi$, and finally, $\bm{l}$ is the unit normal to
the orbital plane. We use the notation $n_\stf{ab} = n_a n_b -
\frac{1}{3} \delta_{ab}$ and $\phi_\stf{ab} = \phi_a \phi_b -
\frac{1}{3} \delta_{ab}$. 

\subsection{Circular motion} 

To specialize to circular orbits we set $\dot{r} = 0$ and $r\dot{\phi}
= v$ in the preceding results. The post-Newtonian equations of motion
imply that 
\begin{equation} 
\omega := \dot{\phi} = \sqrt{\frac{Gm}{r^3}} \Bigl[
1 - \frac{1}{2} (3-\eta) (v/c)^2  + O(c^{-4}) \Bigr], 
\label{7.18}
\end{equation}
where $m = m_1 + m_2$ is total mass and $\eta := m_1 m_2/m^2$ is a
dimensionless reduced mass. As a consequence of Eq.~(\ref{7.18}) we
find that $v = \sqrt{Gm/r} + O(c^{-2})$. Making these substitutions in
Eqs.~(\ref{7.14})--(\ref{7.17}) gives 
\begin{eqnarray} 
{\cal E}_{ab} &=& -\frac{3Gm_2}{r^3} \left\lgroup \biggl[ 
1 - \frac{m_1 + 2m_2}{2m} (v/c)^2 \biggr] n_\stf{ab} 
\right. \nonumber \\ & & \mbox{} \left.
+ (v/c)^2 \phi_\stf{ab} \right\rgroup + O(c^{-4}), 
\label{7.19} \\
{\cal B}_{ab} &=& -\frac{6 G m_2}{r^3} v\, l_{(a} n_{b)} 
+ O(c^{-2}), 
\label{7.20} \\
\dot{A} &=& \frac{ (2m_1 + 3m_2) m_2}{2m^2} v^2 + O(c^{-2}), 
\label{7.21} \\ 
\dot{R}^a &=& -\frac{G m_2}{2r^2} \frac{4m_1+3m_2}{m} v\, 
l^a + O(c^{-2}). 
\label{7.22}
\end{eqnarray} 
For circular orbits we also have $\phi = \omega t$, with $\omega$
given by Eq.~(\ref{7.18}). 

To list the components of the tidal moments it is useful to define 
\begin{eqnarray} 
{\cal E}_0 &:=& \frac{1}{2} ({\cal E}_{11} + {\cal E}_{22}),
\nonumber \\  
{\cal E}_{1c} &:=& {\cal E}_{13},
\nonumber \\ 
{\cal E}_{1s} &:=& {\cal E}_{23}, \label{7.23} \\
{\cal E}_{2c} &:=& \frac{1}{2} ({\cal E}_{11} - {\cal E}_{22}),
\nonumber \\ 
{\cal E}_{2s} &:=& {\cal E}_{12}
\nonumber 
\end{eqnarray}
and 
\begin{eqnarray} 
{\cal B}_0 &:=& \frac{1}{2} ({\cal B}_{11} + {\cal B}_{22}), 
\nonumber \\ 
{\cal B}_{1c} &:=& {\cal B}_{13}, 
\nonumber \\  
{\cal B}_{1s} &:=& {\cal B}_{23}, \label{7.24} \\ 
{\cal B}_{2c} &:=& \frac{1}{2} ({\cal B}_{11} - {\cal B}_{22}), 
\nonumber \\  
{\cal B}_{2s} &:=& {\cal B}_{12}. 
\nonumber
\end{eqnarray}
With the vectorial basis $\bm{n} = (\cos\phi,\sin\phi,0)$, 
$\bm{\phi} = (-\sin\phi,\cos\phi,0)$, and $\bm{l} = (0,0,1)$ we find
that the nonvanishing components of the tidal moments are
\begin{eqnarray} 
{\cal E}_0 &=& -\frac{Gm_2}{2r^3} 
\biggl[ 1 + \frac{m_1}{2m} (v/c)^2 + O(c^{-4}) \biggr], 
\label{7.25} \\ 
{\cal E}_{2c} &=& -\frac{3Gm_2}{2r^3} 
\biggl[ 1 - \frac{3m_1+4m_2}{2m} (v/c)^2 
\nonumber \\ & & \mbox{}
  + O(c^{-4}) \biggr] \cos 2\phi, 
\label{7.26} \\ 
{\cal E}_{2s} &=& -\frac{3Gm_2}{2r^3} 
\biggl[ 1 - \frac{3m_1+4m_2}{2m} (v/c)^2 
\nonumber \\ & & \mbox{}
  + O(c^{-4}) \biggr] \sin 2\phi, 
\label{7.27} \\ 
{\cal B}_{1c} &=& -\frac{3Gm_2}{r^3} v \cos\phi + O(c^{-2}), 
\label{7.28} \\ 
{\cal B}_{1s} &=& -\frac{3Gm_2}{r^3} v \sin\phi + O(c^{-2}). 
\label{7.29}
\end{eqnarray} 
The components ${\cal E}_{1c}$, ${\cal E}_{1s}$, ${\cal B}_0$, 
${\cal B}_{2c}$, and ${\cal B}_{2s}$ all vanish for circular
orbits. These results were already displayed in Sec.~I A; in
Eqs.~(\ref{1.10})--(\ref{1.14}) we used the symbol $b$ (instead of
$r$) for the inter-body distance, $v_{\rm rel}$ (instead of $v$) for
the relative orbital velocity, $m$ (instead of $m_1$) for the
black-hole mass, and $m'$ (instead of $m_2$) for the mass of the
external body.  

\subsection{Tidal moments in the black-hole frame} 

The tidal moments of Eqs.~(\ref{7.25})--(\ref{7.29}) refer to the
barycentric frame. We may express them in the moving frame of the
black hole by invoking the transformation of
Eqs.~(\ref{6.33})--(\ref{6.35}). The transformation involves a switch
from global time $t$ to local time $\bar{t}$, and a rotation of the
Cartesian axes mediated by the vector $\bm{R}(t)$. According to
Eqs.~(\ref{7.21}) the transformation of the time coordinate is given
by  
\begin{equation} 
t = \biggl[ 1 + \frac{(2m_1 + 3m_2) m_2}{2m^2} (v/c)^2 
+ O(c^{-4}) \biggr] \bar{t}.
\label{7.30}
\end{equation} 
And according to Eq.~(\ref{7.22}) we have $c^{-2} R^a 
= -(\Omega \bar{t}) l^a$, with
\begin{equation} 
\Omega := \sqrt{\frac{Gm}{r^3}} 
\frac{(4m_1 + 3m_2)m_2}{2m^2} (v/c)^2 + O(c^{-4}) 
\label{7.31}
\end{equation} 
denoting the precessional angular velocity of the moving frame
relative to the barycentric frame. (This is the rotation of the
coordinate axes, not the rotational motion of the black hole on its
orbit.) The rotation takes place around the $z$-axis, and it is easy
to show that it is effected by the transformation $\phi \to \bar{\phi}  
= \phi - \Omega \bar{t}$. 

Altogether we find that the tidal moments are given by 
\begin{eqnarray} 
\bar{\cal E}_0 &=& -\frac{Gm_2}{2r^3} 
\biggl[ 1 + \frac{m_1}{2m} (v/c)^2 + O(c^{-4}) \biggr], 
\label{7.32} \\ 
\bar{\cal E}_{2c} &=& -\frac{3Gm_2}{2r^3} 
\biggl[ 1 - \frac{3m_1+4m_2}{2m} (v/c)^2 
\nonumber \\ & & \mbox{}
  + O(c^{-4}) \biggr] \cos 2\bar{\phi}, 
\label{7.33} \\ 
\bar{\cal E}_{2s} &=& -\frac{3Gm_2}{2r^3} 
\biggl[ 1 - \frac{3m_1+4m_2}{2m} (v/c)^2 
\nonumber \\ & & \mbox{}
  + O(c^{-4}) \biggr] \sin 2\bar{\phi}, 
\label{7.34} \\ 
\bar{\cal B}_{1c} &=& -\frac{3Gm_2}{r^3} v \cos\bar{\phi}
+ O(c^{-2}),  
\label{7.35} \\ 
\bar{\cal B}_{1s} &=& -\frac{3Gm_2}{r^3} v \sin\bar{\phi}
+ O(c^{-2})
\label{7.36}
\end{eqnarray} 
in the moving frame. Tensorial expressions for $\bar{\cal E}_{ab}$ and
$\bar{\cal B}_{ab}$ can be obtained directly from Eqs.~(\ref{7.19})
and (\ref{7.20}) by making the substitution $\phi \to
\bar{\phi}$. After involvement of Eqs.~(\ref{7.18}), (\ref{7.30}), and
(\ref{7.31}) we find that $\bar{\phi} = \bar{\omega} \bar{t}$, with 
\begin{equation} 
\bar{\omega} = \sqrt{\frac{Gm}{r^3}} \Bigl[
1 - \frac{1}{2} (3+\eta) (v/c)^2 + O(c^{-4}) \Bigr]. 
\label{7.37}
\end{equation} 
This is the angular frequency of the tidal moments as measured in the 
moving frame of the black hole. The transformation from $\omega$ to
$\bar{\omega}$ involves a switch from barycenter time to local proper
time, and a rotation of the local accelerated frame relative to the
global inertial frame. Notice the change in sign in front of 
$\eta := m_1 m_2/m^2$ between Eqs.~(\ref{7.18}) and (\ref{7.37}).    

\subsection{Comparison with Schwarzschild tidal fields} 

The tidal moments of a black hole of small mass $m_1$ moving in the
gravitational field of another black hole of large mass $m_2$ can be
obtained simply by evaluating the components of the Riemann tensor for 
the large black hole; the Riemann tensor is evaluated in the moving
frame of the small black hole. The details of such a computation are
presented in Poisson \cite{poisson:04a}, and in this subsection we
compare our results. Poisson uses different definitions for the
harmonic components of the tidal moments, and his results are
presented in Schwarzschild coordinates. With the conventions adopted 
here, in relativist's units, and in harmonic coordinates, Poisson's
results are   
\begin{eqnarray*} 
\bar{\cal E}_0 &=& -\frac{m_2}{2(r+m_2)^2(r-2m_2)}, \\
\bar{\cal E}_{2c} &=& -\frac{3m_2}{2(r+m_2)^3} \frac{r-m_2}{r-2m_2}
\cos 2\bar{\phi}, \\ 
\bar{\cal E}_{2s} &=& -\frac{3m_2}{2(r+m_2)^3} \frac{r-m_2}{r-2m_2}
\sin 2\bar{\phi}, \\ 
\bar{\cal B}_{1c} &=& -\frac{3m_2^{3/2}}{(r+m_2)^3} 
\frac{\sqrt{r-m_2}}{r-2m_2} \cos\bar{\phi}, \\ 
\bar{\cal B}_{1s} &=& -\frac{3m_2^{3/2}}{(r+m_2)^3} 
\frac{\sqrt{r-m_2}}{r-2m_2} \sin\bar{\phi},
\end{eqnarray*}
where $\bar{\phi} = \bar{\omega} \bar{t}$, with 
\[
\bar{\omega} = \frac{m_2}{(r+m_2)^3}. 
\]
Here, $r$ is the orbital radius of the small black hole (in harmonic
coordinates), and $\bar{t}$ is proper time on the circular orbit. 

It is easy to check that when $(v/c)^2 := m_2/r \ll 1$, the
Schwarzschild expressions reduce to Eqs.~(\ref{7.32})--(\ref{7.37})
when $m_1 \ll m_2$; in this limit $m \simeq m_2$ and $\eta \simeq 0$.   
In their common domain of validity, our results agree with those of
Poisson. 

\section{Geometry of the event horizon} 

In this section we present an application of the results obtained in
Sec.~VII. We examine, in a particular gauge, the intrinsic geometry of
the event horizon of a tidally-deformed black hole. We emphasize
that the discussion presented here is tied to a specific choice of
gauge; a different slicing of the event horizon would produce a
different intrinsic geometry. 

The harmonic coordinates $(\bar{t},\bar{x}^a)$ are singular on the 
black-hole horizon, and an examination of its geometry requires a
change of coordinates. For this purpose we return to the light-cone
coordinates $(v,\rho,\theta^A)$ of Sec.~III B. It is known
\cite{poisson:05} that in the light-cone gauge, the coordinate
description of the event horizon is $\rho = 2M_1 = 2Gm_1/c^2$, the
same as in the unperturbed Schwarzschild geometry. Equations
(\ref{3.10})--(\ref{3.13}) then imply that the induced metric on the
event horizon is given by $g_{AB} = (2M_1)^2 \Omega_{AB} + h_{AB}
+ O(M_1^5 {\cal R}^{-2} {\cal L}^{-1})$, where $\Omega_{AB}$ is the
metric on the unit two-sphere, and  
\begin{equation} 
h_{AB} = -\frac{1}{6} (2M_1)^4 \bigl( \E^{\sf q}_{AB} 
+ \B^{\sf q}_{AB} \bigr) 
\label{8.1}
\end{equation} 
is the tidal perturbation. Here $\E^{\sf q}_{AB}$ and 
$\B^{\sf q}_{AB}$ are the tidal potentials defined in
Eqs.~(\ref{3.16}) and (\ref{3.18}), respectively. As in Sec.~III B, we 
(momentarily) refrain from displaying the factors of $c$ as well as
the overbar. 

To simplify the horizon metric we implement a gauge transformation
generated by the vector field 
\begin{equation} 
\xi_A = -\frac{1}{6} (2M_1)^4 \bigl( \E^{\sf q}_{A} 
+ \B^{\sf q}_{A} \bigr), 
\label{8.2}
\end{equation} 
where $\E^{\sf q}_{A}$ and $\B^{\sf q}_{A}$ are introduced in
Eqs.~(\ref{3.15}) and (\ref{3.17}), respectively. The transformation
changes the metric perturbation according to $h_{AB} \to h'_{AB} =
h_{AB} - D_A \xi_B - D_B \xi_A$, where $D_A$ is the
covariant-derivative operator compatible with $\Omega_{AB}$. Using the
relations $D_A \E^{\sf q}_B = D_B \E^{\sf q}_A 
= \frac{1}{2} \E^{\sf q}_{AB} - \frac{3}{2} \Omega_{AB} \E^{\sf q}$ 
and $D_A \B^{\sf q}_B + D_B \B^{\sf q}_A = \B^{\sf q}_{AB}$, we find
that the new perturbation is given by 
\begin{equation} 
h'_{AB} = -\frac{1}{2} (2M_1)^4 \Omega_{AB} \E^{\sf q}, 
\label{8.3}
\end{equation}
where $\E^{\sf q}$ is defined by Eq.~(\ref{3.14}). 

Reinstating the factors of $c$ and the overbar on the tidal moments
(to emphasize that we are working in the black-hole's comoving frame),
we find that the induced metric on the black-hole horizon is given by 
\begin{equation} 
g_{AB} = (2M_1)^2 \biggl[ 1 - \frac{(2M_1)^2}{2 c^2} 
\bar{\E}_{ab} \Omega^a \Omega^b \biggr] \Omega_{AB} 
+ O\biggl( \frac{M_1^5}{{\cal R}^2 {\cal L}} \biggr)
\label{8.4}
\end{equation}
in this choice of gauge. The line element on the horizon is
\begin{eqnarray} 
ds^2 &=& (2M_1)^2 \biggl[ 1 - \frac{(2M_1)^2}{2 c^2} 
\bar{\E}_{ab} \Omega^a \Omega^b \biggr] 
(d\theta^2 + \sin^2\theta\, d\phi^2)
\nonumber \\ & & \mbox{}
+ O\biggl( \frac{M_1^5}{{\cal R}^2 {\cal L}} \biggr), 
\label{8.5}
\end{eqnarray} 
and $\Omega^a = (\sin\theta\cos\phi, \sin\theta\sin\phi, \cos\theta)$
is the unit radial vector. According to these equations, the area of
each cross-section $v = \mbox{constant}$ of the event horizon is given
by $A = 4\pi (2M_1)^2$, so that $2M_1$ is the radius averaged over
each cross-section. 

The metric of Eqs.~(\ref{8.4}) and (\ref{8.5}) can be reproduced by
embedding a closed two-surface described by  
\begin{equation} 
r = 2M_1 [ 1 + \varepsilon(\theta,\phi) ] 
\label{8.6}
\end{equation} 
in a flat, three-dimensional space charted by spherical coordinates
$(r,\theta,\phi)$. Working consistently to first order in
$\varepsilon$, we find that the metric on this two-surface is given by 
$ds^2 = (2M_1)^2 (1 + 2\varepsilon)(d\theta^2 + \sin^2\theta\,
d\phi^2)$, and this agrees with Eq.~(\ref{8.5}) when 
\begin{equation} 
\varepsilon(\theta,\phi) = -\frac{M_1^2}{c^2} \bar{\E}_{ab} \Omega^a 
\Omega^b. 
\label{8.7}
\end{equation} 
This equation evidently describes a quadrupole deformation of a round
two-sphere of radius $2M_1$. To first order in $\varepsilon$, the
deformation produces no change in area.

So far our considerations have been limited only by the restriction 
$M_1 \ll {\cal R}$, which ensures that the tidal perturbation is
small. As was discussed in Sec.~I A, this restriction includes both
the small-hole and weak-field approximations as limiting cases. If we
now restrict our attention to the weak-field approximation and place
the black hole within a post-Newtonian environment, then the tidal
moments $\bar{\E}_{ab}$ can be imported from Sec.~VI D and substituted
within Eq.~(\ref{8.7}). To illustrate this, we restrict our attention
further to the situation examined in Sec.~VII C, in which the black
hole is a member of a two-body system in circular motion. The relevant
tidal moments are listed in Eqs.~(\ref{7.32})--(\ref{7.34}), and when
these are inserted within Eq.~(\ref{8.7}), we obtain  
\begin{eqnarray} 
\varepsilon &=& \frac{M_1^2 M_2}{2 b^3} \biggl[ 1 + \frac{M_1}{2M}
(v_{\rm rel}/c)^2 + O(c^{-4}) \biggr] (1 - 3\cos^2\theta) 
\nonumber \\ & & \mbox{}
+ \frac{3 M_1^2 M_2}{2 b^3} \biggl[ 1 - \frac{3M_1 + 4M_2}{2M}
(v_{\rm rel}/c)^2 
\nonumber \\ & & \mbox{}
+ O(c^{-4}) \biggr] \sin^2\theta \cos 2\psi.   
\label{8.8}
\end{eqnarray}   
Here $M_1 := G m_1/c^2$ is the black hole's gravitational radius, and
$M_2 := G m_2/c^2$ measures the mass of the companion body. We
use the same notation as in Sec.~I B: $b$ is the separation between
the two bodies (in harmonic coordinates), $M := M_1 + M_2$ is a
measure of the total mass within the system, and $v_{\rm rel} =
c\sqrt{M/b}$ is the relative orbital velocity. The symbol $\psi$
stands for $\phi - \bar{\omega} v$, where $v$ is the advanced-time
coordinate on the event horizon and $\bar{\omega}$ is the orbital
frequency of Eq.~(\ref{7.37}).

Equation (\ref{8.8}) implies that the event horizon is bulging along
an axis directed toward the orbiting body. To see this clearly, we
calculate from the metric the circumference of a line of longitude
$\psi = \mbox{constant}$, and we obtain 
\begin{eqnarray} 
C_{\rm l} &=& 2\pi (2M_1) \biggl\{ 1 - 
\frac{M_1^2 M_2}{4 b^3} \biggl[ 1 + \frac{M_1}{2M} 
(v_{\rm rel}/c)^2 
+ O(c^{-4}) \biggr]
\nonumber \\ & & \mbox{}
+ \frac{3 M_1^2 M_2}{4 b^3} \biggl[ 1 - \frac{3M_1 + 4M_2}{2M}
(v_{\rm rel}/c)^2 
\nonumber \\ & & \mbox{}
+ O(c^{-4}) \biggr] \cos 2\psi \biggr\}. 
\label{8.9}
\end{eqnarray} 
This equation reveals that the circumference is largest (stretched)
when $\psi = \{0,\pi\}$ and smallest (squeezed) when $\psi =  
\{\frac{\pi}{2},\frac{3\pi}{2}\}$. We also calculate the circumference
of the equator (at $\theta = \frac{\pi}{2}$) and obtain 
\begin{eqnarray} 
C_{\rm e} &=&  2\pi (2M_1) \biggl\{ 1 + 
\frac{M_1^2 M_2}{2 b^3} \biggl[ 1 + \frac{M_1}{2M} 
(v_{\rm rel}/c)^2 
\nonumber \\ & & \mbox{}
+ O(c^{-4}) \biggr] \biggr\}. 
\label{8.10}
\end{eqnarray} 
This equation also reveals a bulging of the horizon at the equator.  

\section{Tidal heating} 

In this final section, we present another application of the results 
obtained in Sec.~VII. We calculate the tidal heating of a black hole
of mass $m_1$ placed in a post-Newtonian tidal environment created by
an external body of mass $m_2$. For simplicity, we restrict our
attention to circular motion. These results, unlike those presented in 
Sec.~VIII, are gauge-invariant. In Sec.~IX A we calculate the tidal
heating of a nonrotating black hole, and in Sec.~IX B we examine the
case of a rotating black hole. The foundations for this calculation
are given in Ref.~\cite{poisson:04d}.  

\subsection{Nonrotating black hole} 

The rate at which a black hole of mass $m_1$ acquires mass by tidal
heating is given by \cite{poisson:04d}
\begin{equation} 
G \dot{m}_1 = \frac{16}{45} \frac{(Gm_1)^6}{c^{15}} \biggl[ 
\dot{\bar{\cal E}}_{ab} \dot{\bar{\cal E}}^{ab} 
+ \frac{1}{c^2} \dot{\bar{\cal B}}_{ab} \dot{\bar{\cal B}}^{ab}
+ O(c^{-4}) \biggr], 
\label{9.1} 
\end{equation} 
in which an overdot indicates differentiation with respect to
$\bar{t}$. This equation excludes contributions from octupole and
higher-order tidal moments --- see Ref.~\cite{poisson:05}. It is easy
to show, however, that these contributions occur at order $c^{-4}$
(and smaller) relative to the dominant, quadrupole term; they are
therefore included in the neglected terms of Eq.~(\ref{9.1}). 

According to our results in Sec.~VII D, the tidal moments are given by   
\begin{eqnarray} 
\bar{{\cal E}}_{ab} &=& -\frac{3Gm_2}{r^3} \left\lgroup \biggl[ 
1 - \frac{m_1 + 2m_2}{2m} (v/c)^2 \biggr] \bar{n}_\stf{ab} 
\right. \nonumber \\ & & \mbox{} \left.
+ (v/c)^2 \bar{\phi}_\stf{ab} \right\rgroup + O(c^{-4}), 
\label{9.2} \\
\bar{\cal B}_{ab} &=& -\frac{6 G m_2}{r^3} v\, 
\bar{l}_{(a} \bar{n}_{b)}  + O(c^{-2})
\label{9.3}
\end{eqnarray}
in the black-hole frame. Here $\bar{n}^a 
= (\cos\bar{\phi},\sin\bar{\phi},0)$, $\bar{\phi}^a 
= (-\sin\bar{\phi}, \cos\bar{\phi}, 0)$, and $\bar{l}^a = (0,0,1)$, 
with $\bar{\phi} = \bar{\omega} \bar{t}$; the angular frequency 
$\bar{\omega}$ is displayed in Eq.~(\ref{7.37}). The time derivatives 
of the basis vectors are given by $\dot{\bar{n}}_a 
= \bar{\omega} \bar{\phi}_a$, $\dot{\bar{\phi}}_a 
= -\bar{\omega} \bar{n}_a$, and $\dot{\bar{l}}_a = 0$; this gives 
rise to $\dot{\bar{n}}_\stf{ab} = 2\bar{\omega}\, \bar{n}_{(a}
\bar{\phi}_{b)}$ and $\dot{\bar{\phi}}_\stf{ab} = -2\bar{\omega}\, 
\bar{n}_{(a} \bar{\phi}_{b)}$. 

Evaluating Eq.~(\ref{9.1}) from Eqs.~(\ref{9.2}) and (\ref{9.3})
produces 
\begin{eqnarray} 
G\dot{m}_1 &=& \frac{32}{5 c^{15}} \frac{m_1^6 m_2^2}{m^8} 
\Bigl(\frac{Gm}{r}\Bigr)^9 \biggl[ 
1 
\nonumber \\ & & \mbox{} \hspace*{-35pt}
- \frac{5m_1^2 + 12m_1 m_2 + 6m_2^2}{m^2} (v/c)^2 
+ O(c^{-4}) \biggr], 
\label{9.4} 
\end{eqnarray}
where $m = m_1 + m_2$ is the total mass, $r$ is the orbital separation
(in harmonic coordinates), and $v = \sqrt{Gm/r}$ is the relative
orbital velocity. The rate at which the tidal coupling increases the
black hole's angular momentum can next be obtained from the
rigid-rotation relation $\dot{m}_1 c^2 = \bar{\omega}
\dot{J}_1$. Equation (\ref{9.4}) was already displayed in Sec.~I B; in
Eq.~(\ref{1.17}) we used the symbol $b$ (instead of $r$) for the
orbital separation, $v_{\rm rel}$ (instead of $v$) for the relative
orbital velocity, $m$ (instead of $m_1$) for the black-hole mass, and
$m'$ (instead of $m_2$) for the mass of the external body.    

Equation (\ref{9.4}) can be compared with the result obtained by
Poisson \cite{poisson:04d} for a black hole of small mass $m_1$ moving
in the field of another black hole of large mass $m_2$. In geometrized
units, and in harmonic coordinates, Poisson's result is 
\[
\dot{m}_1 = \frac{32}{5} \Bigl( \frac{m_1}{m_2} \Bigr)^6 
\Bigl( \frac{m_2}{r} \Bigr)^9 
\frac{1 - m_2/r}{(1+m_2/r)^9 (1-2m_2/r)^2}. 
\]
When $m_2/r = (v/c)^2$ is small the relativistic factor becomes
$1 - 6(v/c)^2 + O(c^{-4})$, and this expression agrees with
Eq.~(\ref{9.4}) when $m_1 \ll m_2$.   
 
\subsection{Rotating black hole} 
 
We next calculate the tidal heating of a rotating black hole, assuming
that the tidal fields are not affected (at first post-Newtonian order)
when the nonrotating black hole is replaced by a rapidly rotating
hole. The rate at which the hole's angular momentum is increased by
the tidal coupling is given by \cite{poisson:04d}  
\begin{eqnarray} 
G \dot{J}_1 &=& -\frac{2}{45} \frac{(Gm_1)^5}{c^{10}} \chi 
\Bigl[ 8(1+3\chi^2) (E_1 + c^{-2} B_1) 
\nonumber \\ & & \mbox{}
- 3(4+17\chi^2) (E_2 + c^{-2} B_2) 
\nonumber \\ & & \mbox{}
+ 15\chi^2 (E_3 + c^{-2} B_3) + O(c^{-4}) \Bigr], 
\label{9.5} 
\end{eqnarray}
where $\chi := cJ_1/(G m_1^2)$ is a dimensionless angular-momentum
parameter that ranges between 0 and 1, and 
$E_1 := \bar{\cal E}_{ab} \bar{\cal E}^{ab}$,  
$E_2 := (\bar{\cal E}_{ab} s^b) (\bar{\cal E}^a_{\ c} s^c)$, 
$E_3 := (\bar{\cal E}_{ab} s^a s^b)^2$, 
$B_1 := \bar{\cal B}_{ab} \bar{\cal B}^{ab}$, 
$B_2 := (\bar{\cal B}_{ab} s^b) (\bar{\cal B}^a_{\ c} s^c)$, 
$B_3 := (\bar{\cal B}_{ab} s^a s^b)^2$. Here, the unit vector $s^a$
points in the direction of the hole's spin angular-momentum vector, so 
that $\bm{J}_1 = J_1 \bm{s}$. In this application, the spin and
orbital angular momenta are aligned or antialigned, so that $s^a =
\epsilon l^a$ with $\epsilon = \pm 1$.  

Evaluation of Eq.~(\ref{9.5}) produces 
\begin{eqnarray} 
G\dot{J}_1 &=& -\frac{8}{5} \chi (1+3\chi^2) 
\frac{m_1^5 m_2^2}{m^7} \frac{(Gm)^7}{c^{10} r^6} 
\biggl[ 1 
\nonumber \\ & & \mbox{}
- \biggl( \frac{8+39\chi^2}{4+12\chi^2} \frac{m_1}{m} 
+ \frac{12+51\chi^2}{4+12\chi^2} \frac{m_2}{m} \biggr) (v/c)^2
\nonumber \\ & & \mbox{}
+ O(c^{-4}) \biggr],
\label{9.6}
\end{eqnarray} 
where $m = m_1 + m_2$ is the total mass, $r$ the orbital separation 
(in harmonic coordinates), and $v = \sqrt{Gm/r}$ is the orbital
velocity. The rate at which the black-hole mass changes as a result
of tidal heating can next be obtained from the rigid-rotation relation
$\dot{m}_1 c^2 = \bar{\omega} \dot{J}_1$. With Eq.~(\ref{7.37}) we get  
\begin{eqnarray} 
G\dot{m}_1 &=& -\frac{8\epsilon}{5 c^{12}} \chi (1+3\chi^2)  
\frac{m_1^5 m_2^2}{m^7} \Bigl( \frac{Gm}{r} \Bigr)^{15/2}  
\biggl[ 1 
\nonumber \\ & & \mbox{}
- \biggl( \frac{14+57\chi^2}{4+12\chi^2} \frac{m_1^2}{m^2}  
+ \frac{34+132\chi^2}{4+12\chi^2} \frac{m_1 m_2}{m^2} 
\nonumber \\ & & \mbox{}
+ \frac{18+69\chi^2}{4+12\chi^2} \frac{m_2^2}{m^2} \biggr) (v/c)^2 
+ O(c^{-4}) \biggr],
\label{9.7}
\end{eqnarray} 
where the parameter $\epsilon = \pm 1$ was previously defined by the
relation $s^a = \epsilon l^a$. Thus, the black hole {\it loses mass}
when the orbital motion proceeds in the same direction as the spinning 
motion ($\epsilon = 1$), and it {\it gains mass} when the 
orbital motion proceeds in the opposite direction ($\epsilon =
-1$). In each case the orbital motion is slower than the spinning
motion, and the black hole always loses angular momentum.  

Equation (\ref{9.7}) can be compared with the result obtained by
Poisson \cite{poisson:04d} for a rotating black hole of small mass
$m_1$ moving in the field of a nonrotating black hole of large mass
$m_2$. In geometrized units, and in harmonic coordinates, Poisson's
result is 
\begin{eqnarray*}
\dot{m}_1 &=& -\frac{8\epsilon}{5} \chi (1+3\chi^2) 
\Bigl(\frac{m_1}{m_2}\Bigr)^5 \Bigl(\frac{m_2}{r}\Bigr)^{15/2} 
\nonumber \\ & & \mbox{} \times 
\frac{(1-m_2/r)(1 - \frac{15\chi^2}{4+12\chi^2} m_2/r)}
   {(1+m_2/r)^{15/2} (1-2m_2/r)^2}.
\end{eqnarray*}
When $m_2/r = (v/c)^2$ is small the relativistic factor becomes 
\[
1 - \frac{18+69\chi^2}{4+12\chi^2} (v/c)^2 + O(c^{-4}),
\]
and this expression agrees with Eq.~(\ref{9.7}) when $m_1 \ll
m_2$. The same conclusion holds when the small black hole moves in the
field of a large rotating black hole. In this situation, the error
term in the previous expression is of order $c^{-3}$ instead of order 
$c^{-4}$.   

\begin{acknowledgments} 
This work was supported by the Natural Sciences and Engineering
Research Council of Canada. We thank Steve Detweiler, Harald
Pfeiffer, \'Etienne Racine, and Igor Vlasov for useful conversations.       
\end{acknowledgments} 

\bibliography{../bib/master}
\end{document}